\documentclass[onecolumn, 11pt]{IEEEtran}
\usepackage{epsfig}
\usepackage{amssymb,amsfonts}
\usepackage{amsmath}
\usepackage{setspace}
\usepackage{lscape}

\newcommand{\Real}{\mathbb{R}}
 \newcommand{\Complex}{\mathbb{C}}

\def \prob {\mathbb{P}}

\def \bW{\mathbf{W}}

\def \bX{\mathbf{X}}
\def \bU{\mathbf{U}}

\def \bB{\mathbf{B}}
\def \bphi{\mathbf{\phi}}
\def \bPhi{\mathbf{\Phi}}

\def \ex {\mathbf{E}}

\def \bY{\mathbf{Y}}

\def \bX{\mathbf{X}}

\def \bv{\mathbf{v}}

\def \bA{\mathbf{A}}
\def \bPhi{\mathbf{\Phi}}

\def \b1 {\mathbf{1}}

\def \bAt{\mathbf{A}_{\Theta}}
\def \bB{\mathbf{B}}

\newtheorem{remark}{Remark}[section]
\newtheorem{defn}{\indent \bf Definition}[section]

\begin{document}

\title{Joint multi-mode dispersion extraction in Fourier and space time domains}

\author{{Sandip~Bose,~\IEEEmembership{Member,~IEEE,}
        Shuchin~Aeron,~\IEEEmembership{Member,~IEEE,}
        and~Henri-Pierre~Valero}
\thanks{Sandip Bose is with the Math and Modeling Department, Schlumberger Doll Research (SDR), Cambridge, MA, USA. e-mail: bose1@slb.com}
\thanks{Shuchin Aeron is with the Dept. of Electrical and Computer Engineering (ECE) at Tufts University, Medford, MA, USA. e-mail:shuchin@ece.tufts.edu }
\thanks{Henri-Pierre Valero is with SKK, Japan. e-mail:hvalero@slb.com}}

\maketitle

\begin{abstract}
In this paper we present a novel broadband approach for the extraction of dispersion curves of multiple time frequency overlapped dispersive modes such as in borehole acoustic data.  The new approach works jointly in the Fourier and space time domains and, in contrast to existing space time  approaches that mainly work for time frequency separated signals, efficiently handles multiple signals with significant time frequency overlap. The  proposed method begins by exploiting the slowness (phase and group) and time location estimates based on frequency-wavenumber (f-k) domain sparsity penalized broadband  dispersion extraction method as presented in \cite{AeronTSP2011}. In this context we first present a Cramer Rao Bound (CRB) analysis for slowness estimation in the (f-k) domain and show that for the f-k domain broadband processing, group slowness estimates have more variance than the phase slowness estimates and time location estimates. In order to improve the group slowness estimates we exploit the time compactness property of the modes to effectively represent the  data as a linear superposition of time compact space time propagators parameterized by the phase and group slowness. A linear least squares estimation algorithm in the space time domain is then used to obtain improved group slowness estimates. The performance of the method is demonstrated  on real borehole acoustic data sets.
\end{abstract}

\begin{keywords}
Dispersion extraction, sparsity, continuous wavelet transform, space time methods
\end{keywords}

\section{Introduction}
\label{sec:intro}

Acoustic data collected by borehole logging tools is dominated by wave guided propagating modes.  Such modes are usually dispersive, that is, each mode propagates at a slowness (inverse of velocity) that varies as a function of frequency. This dispersion or frequency dependent variation of phase slowness is an important characteristic of such modes and is now understood as conveying useful information about the rock formations traversing the borehole \cite{Paillet,Sinha94,Sinha98}. Existing techniques for dispersion extraction employ a frequency by frequency processing or so called narrowband processing and extract the wavenumber at each frequency using uniform linear array (ULA) processing techniques such as the Prony \cite{McClellan87} and matrix pencil \cite{Ekstrom95} methods for analyzing exponentials.  A number of other approaches for dispersion analysis have been proposed such as phase minimization or coherency maximization \cite{NolteHuang97},  or slowness frequency coherence based methods \cite{Tang04,RamaRao05}.  Although adequate for analyzing strong signals in low to moderate noise situations, i.e with favorable signal to noise ratio (SNR),  these narrowband approaches fail to identify all the modes of interest, especially the weaker ones, and may not be  robust and accurate enough for quantitative interpretation even for such strong signals
due to scatter and aliasing in the results.   This is particularly an issue  in logging while drilling (LWD) scenarios where the SNR could be low and there could be significant interference and model deviation due to the drilling process and the presence of a heavy drill collar moving w.r.t. the borehole.  Wireline logging scenarios involving more complex physics such as in cased hole or deviated wells may also prove challenging for such methods.  Other array processing methods such as ESPRIT and MUSIC, \cite{Roy89,Viberg89} can be employed here but are sensitive to the choice of \emph{model order} , \cite[Chapter 16]{Viberg89}, and the knowledge of signal subspace. Schemes such as in \cite{Kailath85,Viberg89} based on minimum description length (MDL), \cite{Rissanen78} have been proposed to estimate the signal subspace and the model order. However these methods work fine under some stationarity assumptions on the signal and are not applicable to borehole acoustic logging where the signal is transient and highly non-stationary.  Thus narrowband methods are inherently limited for the quantitative extraction of modal dispersion curves.   

Therefore there is a need for an efficient \emph{broadband dispersion extraction} algorithm that does not require assumptions of stationarity, and is robust to noise and presence of strong interference due to the coherent processing of the signal over a frequency band. A broadband approach for extraction of a \emph{single mode} dispersion was proposed in \cite{Wang04} which is limited to a particular type of borehole mode namely the lowest order flexural mode of an open borehole or one where the effect of the tool is well characterized.   A more general approach for extracting dispersion of a single mode without reference to a physical model or restriction of the type of mode was proposed in \cite{Pedersen05} and \cite{AeronICASSP08}. There the continuous wavelet transform (CWT) of the array data was employed along with a parameterization of the wavenumber dispersion in the $f-k$ domain in terms of a local  first order linear approximation at each frequency.  It was shown that under such an approximation around the center frequency at any given scale, the shifts in the CWT coefficients at that scale across the receivers is proportional to the group slowness and the phase change is dependent on both phase and group slowness. This property was exploited in \cite{AeronICASSP08} to estimate the dispersion parameters in the CWT domain using a modification of the Radon transform called the exponentially projected Radon transform (EPRT). Although broadband in nature, with good performance for multiple modes in situations where there is significant time separation, these methods have poor performance for multiple modes with significant time overlap in the modes. Similar ideas that exploit the time frequency localization have been considered before, e.g., the energy reassignment method of \cite{Auger95} which has been applied in \cite{Prossner99} for analysis of dispersion of Lamb waves. Again these methods are suitable when the time-frequency separation of modes is good but fails otherwise. In \cite{Hsu1990} a broadband method based on linear approximation of the dispersion curves in the $(f-k)$ domain was proposed. However this method is still based on some stationarity assumptions on the signal which, as pointed out before, does not hold true for real signals obtained with pulsed excitation in downhole logging tools. 

In this context in \cite{AeronTSP2011} the authors proposed a novel sparsity penalized broadband dispersion extraction method in the $f-k$ domain which has been shown to be robust to the presence of time overlapped modes. The key ingredient there was to first propose an overcomplete dictionary of broadband propagators for representation of array data followed by exploiting the \emph{sparsity} in the number of modes, manifested as \emph{simultaneous sparsity}, see \cite{Tropp}, in the representation coefficients at each frequency scale. In this framework as shown in \cite{AeronTSP2011}, a sparsity penalized recovery algorithm leads to dramatic improvements in performance under heavy noise conditions. However the approach is not at par with semblance based processing techniques such as the space time coherence (STC) processing of non-dispersive waves, \cite{Kimball84},  or of dispersive modes such as EPRT,  \cite{AeronICASSP08}, which are more reliable for direct estimation of \emph{group slowness}. The main reason is that unlike STC based techniques such as EPRT, we don't exploit the \emph{local spatio-temporal coherence} of modes in space-time domain. This leads to increased variance in the group slowness estimates from $f-k$ domain processing as shown by a Cramer-Rao Bound (CRB) analysis in the Appendix of this paper. To overcome this drawback, in this paper we combine the two techniques of \cite{AeronTSP2011} and \cite{AeronICASSP08} and propose a joint $f-k$ and space-time domain multimode dispersion extraction method. 

The organization of the paper is as follows.  The next section~\ref{sec:problem_setup} introduces the problem setup and formulation of the approach using the CWT and reviews the use of the broadband $f-k$ processing for the dispersion estimation. Section~\ref{sec:New_Approach} presents the proposed space-time approach to refine the dispersion estimates.  Section~\ref{sec:Real_Data} covers application of the new approach to a diverse series of real data examples indicating the improvement in the quality of the estimates. Finally some performance analysis on  is included in the appendix~\ref{sec:CRB}  after the conclusion.

\section{Problem set-up and broadband formulation}
\label{sec:problem_setup}
\subsection{Preliminaries}
\begin{figure}[t]
\centering \makebox[0in]{
    \begin{tabular}{c}
    \includegraphics[height= 3 in, width = 5.5in]{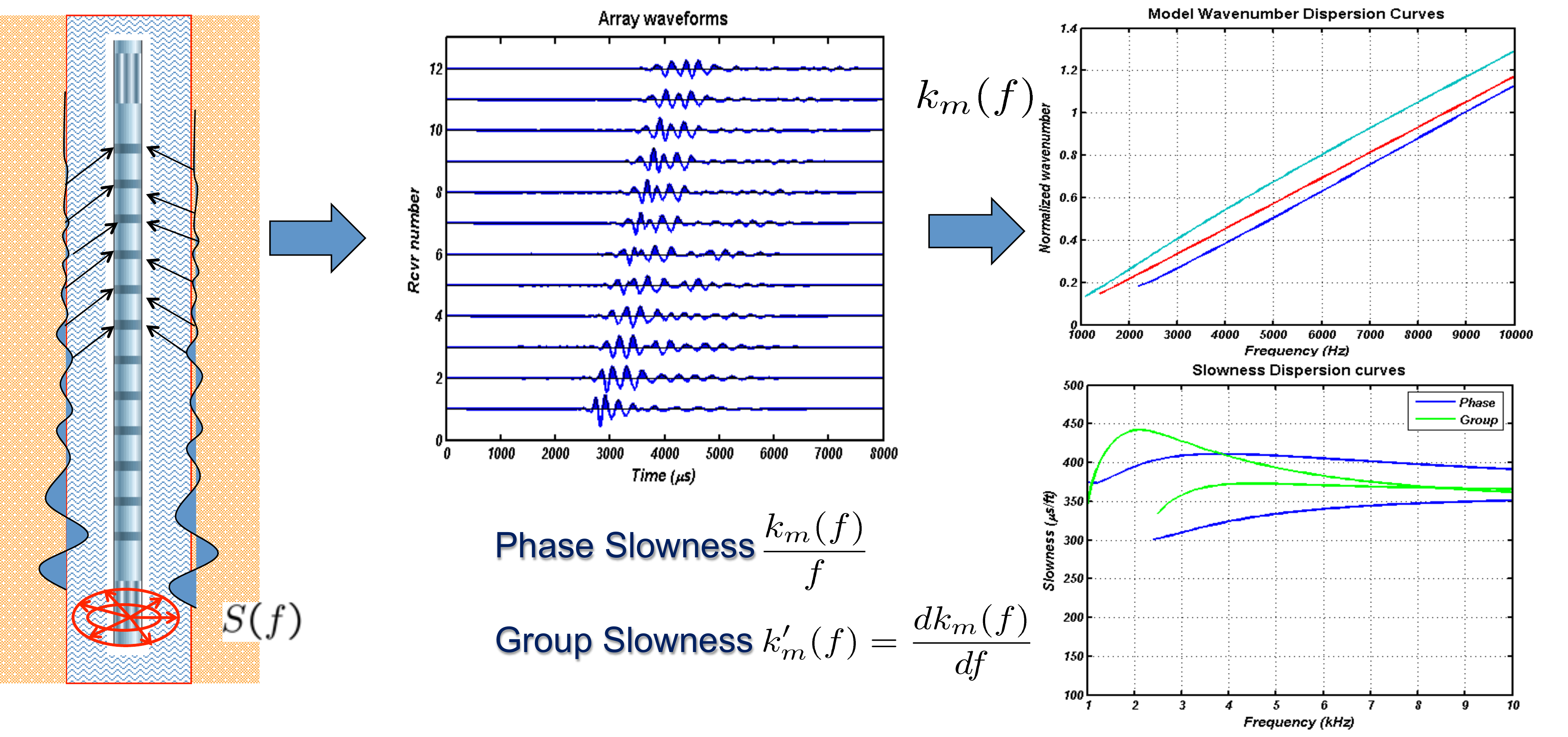}
      \end{tabular}}
  \caption{Schematic depicting the dispersion extraction problem for the case of borehole acoustic data. The figure illustrates an acoustic tool in the borehole generating energy that propagates as a sum of borehole guided modes each governed by a dispersion curve. The task is to estimate these dispersion curves given the waveforms received by an array of receivers on the tool as shown.}
  \label{fig:problem_setup}
\end{figure}

Acoustic waves are excited in the borehole by firing a source which produces \emph{modes} with the borehole as a waveguide. A schematic is shown in Figure~\ref{fig:problem_setup}. The relation between the received waveforms and the frequency-wavenumber $(f-k)$ response of the borehole to the source excitation is captured via the following equation,
\begin{align}
\label{eq_model} s_l(t) = \int_{0}^{\infty}\int_{0}^{\infty} S(f) Q(k,f) e^{i2\pi f t}e^{-i2\pi k z_l} df dk
\end{align}
for $l = 1,2,...,L$, where $s_l(t)$ denotes the pressure at time $t$ at the $l$-th receiver located at a distance $z_l$ from the source; $S(f)$ is the source spectrum and $Q(k,f)$ is the wavenumber-frequency response of
the borehole. Typically the data is acquired in presence of noise (environmental and receiver noise) and interference which we collectively denote by $w_l(t)$. Then the noisy observations at the set of receivers, $y_l(t), l=1,\ldots,L$, can be written as,
\begin{align}
y_l(t) & =  s_l(t) + w_l(t) \label{eq:model_noise}
\end{align}
It has been shown that the complex integral in the wavenumber ($k$) domain in Equation(\ref{eq_model}) can be approximated by the contribution due to the residues of the poles of the system response,
\cite{Paillet,Kurkjian85}. Specifically,
\begin{eqnarray}
\label{eq:approx} \int_{0}^{\infty} Q(k,f) e^{-ikz_l} dk \sim \sum_{m=1}^{M(f)} q_m(f) e^{- (i2\pi k_m(f) + a_m(f))z_l}
\end{eqnarray}
where $q_m(f)$ is the residue of the $m$-th pole and respectively where  $k_m(f)$ and $a_m(f)$ are the corresponding real and imaginary parts of the poles and can be interpreted as the wavenumber and attenuation respectively as functions of frequency. This means that we have
\begin{align}
s_l(t) & =  \int_{0}^{\infty} \sum_{m=1}^{M(f)} S_m(f) e^{- (i2\pi k_m(f) + a_m(f))z_l} e^{i2\pi f t} df \label{eq:exp_model}
\end{align}
where $S_{m}(f) = S(f) q_m(f) $.  
In theory the number of modes can be infinite but in practice only a few are significant, i.e. the model order $M(f)$ is finite and is small. In the scope of this work,  we will assume that $M(f) = M$ for all the frequencies of interest.
Under this signal model, the data acquired at each frequency across the receivers can be written as,
\begin{align}
\label{eq:narrowband} \underset{\bY(f)}{\underbrace{\begin{bmatrix}
  Y_{1}(f) \\
  Y_2(f)\\
\vdots\\
  \vdots\\ 
  Y_{L}(f) \\
\end{bmatrix}}} = \begin{bmatrix}
                  e^{-(i2\pi k_1(f) + a_1(f))z_1} & \hdots & e^{-(i2\pi k_M(f) + a_M(f)) z_1} \\
                  e^{-(i2\pi k_1(f) + a_1(f))z_2} & \hdots & e^{-(i2\pi k_M(f) + a_M(f)) z_2} \\
                  \vdots & \hdots & \vdots\\
                  e^{-(i2\pi k_1(f) + a_1(f)) z_L} & \hdots & e^{-(i2\pi k_M(f) + a_M(f)) z_L} \\
                \end{bmatrix}\underset{\mathbf{S}(f)}{\underbrace{\begin{bmatrix}
                                    S_{1}(f) \\
                                    S_{2}(f)\\
                                    \vdots\\
                                    \vdots\\
                                    S_{M}(f) \\
                             \end{bmatrix}}} + \underset{\mathbf{W}(f)}{\underbrace{\begin{bmatrix}
                                                W_{1}(f) \\
W_2(f)\\
\vdots\\
                                                \vdots\\
                                                W_{L}(f) \\
                                            \end{bmatrix}}}
\end{align}
In other words the data at each frequency is modelled as a superposition of $M$ exponentials sampled with respect to the receiver locations $z_1,..,z_L$. We refer to the above system of equations as corresponding to a sum of exponentials model at each frequency.

We note that the modal dispersion is completely characterized in the $(f-k)$ domain in terms of the wavenumber response $k(f)$ as a function of frequency.   This is often expressed in terms of related physical quantities - the slowness of the phase wavefront or the phase slowness, $s^{\phi}(f)$, and the group slowness, $s^g(f)$ ,  which is the reciprocal of the energy transport velocity, both considered as functions of frerquency.  These are related to the wavenumber and each other as follows:
\begin{align} 
s^{\phi}(f) = \frac{k(f)}{f}  \ \ \ , \ \  s^g(f) = \frac{dk}{df} \ \ \ , \ \ \  s^g(f) = s^{\phi}(f) + f \frac{d s^{\phi}}{df} 
\end{align}

The dispersion extraction problem therefore consists in recovering the frequency dependent phase and group slownesses of the $m^{th}$ mode, $\{s^{\phi}_m(f),s^g_m(f)\}$   from the received data for $m = 1,\ldots,M$ significant propagating modes.  
This is illustrated in Figure~\ref{fig:problem_setup} where we observe two-three significant modes with a typical array acquisition with a borehole sonic tool such as described in \cite{SonicScan06}. Since we are going to use continuous wavelet transform (CWT) for dispersion extraction, in the next section we will provide a brief overview of the broadband dispersion extraction framework in the CWT domain and point out the time-frequency and space time properties of the acoustic modes. 

\begin{remark}
In the following we will assume that the data is sampled temporally at at-least the Nyquist rate and that the total number of time samples is (say) $T$. Therefore we have a discrete time data $\bY \in \Real^{L \times T}$ from which we want to extract the dispersion characteristics at the discrete set of DFT frequencies corresponding to the sampled system. 
\end{remark}

\subsection{Broadband dispersion extraction set-up in the CWT domain}
Recall that the continuous wavelet transform (CWT) of a signal is obtained from  the scalar product of this signal with elements of a wavelet family generated by translations and dilations of a mother wavelet, $g$ by $T^{b}D^{a}[g(t)] = \frac{1}{\sqrt{a}} g(\frac{t -b }{a})$. 
The CWT coefficients $C^s(a,b)$  of the signal $s(t)$ at the  scale (dilation factor) $a$ and shift (translation) $b$ is then given by
\begin{align}
C^s(a,b) & = \frac{1}{\sqrt{a}} \int s(t) g^{*}(\frac{t-b}{a}) dt\\
& =\int_{-\infty}^{\infty} G^*(af) e^{i2\pi b f} S(f) df
\end{align}
where $G(f)$ is the Fourier transform of the analyzing (mother) wavelet $g(t)$ and $S(f)$ is the Fourier transform of the signal
being analyzed. The analyzing wavelet $g(t)$ is chosen to satisfy some admissibility condition, see
\cite{Grossmann84,Grossmann89} for details. The CWT generates a time-scale representation of a signal which indicates the frequency content of the signal as a function of time.  Such a representation is useful for dispersion analysis as explained below. 

\begin{figure*}
\centering \makebox[0in]{
    \begin{tabular}{c}
      \includegraphics[height= 2.5 in]{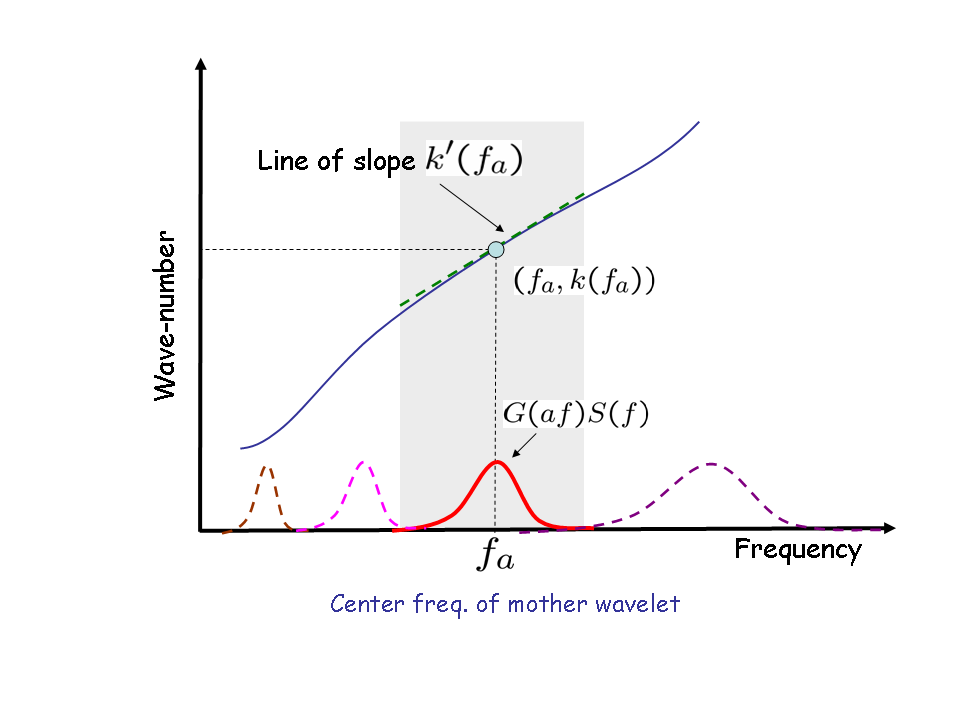}
      \end{tabular}}
  \caption{Schematic showing the first order Taylor series approximation to the dispersion curve in a band around the center
  frequency $f_a$ of the mother wavelet at scale $a$. }
  \label{fig:Taylor_fk_cwt}
\end{figure*}

\begin{figure*}
\centering \makebox[0in]{
    \begin{tabular}{cc}
      \includegraphics[height= 2.5 in]{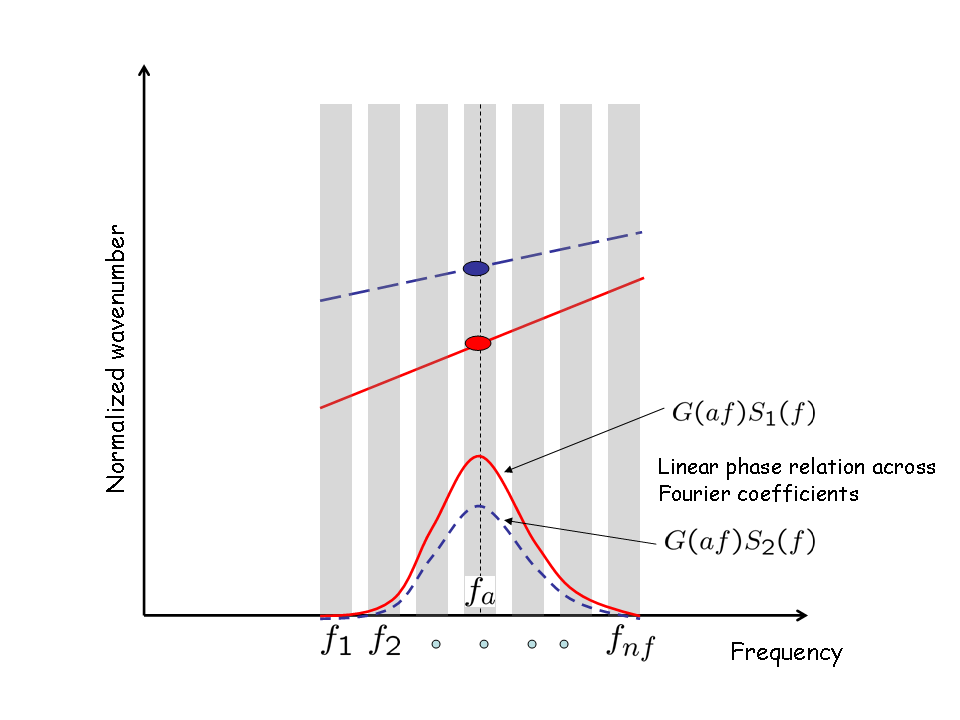} & \includegraphics[height= 2.5 in]{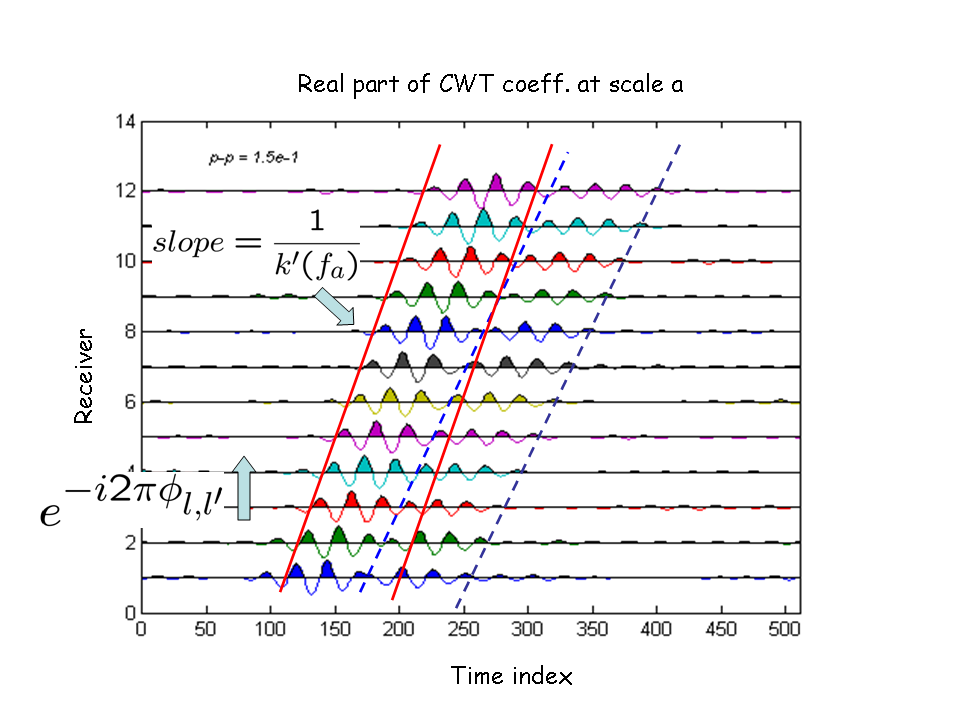}\\
(a) & (b)
      \end{tabular}}
  \caption{Combining the space-time and frequency-wavenumber processing for dispersion extraction using the CWT. Figure depicts the (a) $(f-k)$ domain and (b) time domain characteristics of the CWT coefficients in terms of the propagation parameters at a given scale. The left hand plot shows the linearized wavenumber dispersion curves of two overlapping modes in a frequency band corresponding to the frequency support of the CWT coefficients at scale $a$.  The spectra of the CWT coefficients  for each mode are also shown illustrating the support in the band around the center frequency $f_a$.  The right hand plot shows the real part of the corresponding CWT complex coefficients  indicating the envelope moveout given by the group slowness (frequency derivative of wavenumber) at $f_a$ and an inter-receiver phase shift given by the difference between the phase and group slowness at $f_a$.  Both representations are used for robustly extracting the phase and group slowness estimates at various $f_a$. }
  \label{fig:joint_schematic}
\end{figure*}

Consider the case of a mode $m$ with spectrum $S_m(f)$. Then under the exponential model, the CWT at scale $a$ and time shift $b$ of the received waveform at the $l$-th receiver is given by
\begin{align}
C_{l}^{m}(a,b) = \int_{-\infty}^{\infty} G^{*}(af) S_m(f) e^{i2\pi(b f - z_l k_m(f))} df
\end{align}.
Equivalently by taking the Fourier transform with respect to $b$, one can write, 
\begin{align}
C_{l}^{m}(a,f) = G^{*}(af) S_m(f) e^{-i2\pi z_l k_m(f))} df
\end{align}.
Assuming that a first order Taylor series approximation holds true for the wavenumber in the effective frequency band,  $F_a$, around the center frequency $f_a$ of the analyzing wavelet, i.e. under the approximation (see Figure~\ref {fig:Taylor_fk_cwt}), 
\begin{align}
k_m(f)  &  \approx  k_m(f_a) + k_m^{'} (f_a) (f - f_a) ,  \, \, \, f \in F_a \\ 
C_{l}^{m}(a,f) & \approx G^{*}(af) S_m(f) e^{- i2\pi (k_m(f_a) + k_{m}^{'}(f_a)(f - f_a)) z_l}, \,\,\, f \in F_a
\end{align}
 the CWT coefficients at scale $a$ for the $m^{th}$ mode waveform at the $l$-th receiver can be expressed in terms of a corresponding quantity at a reference ($0^{th}$) receiver,
\begin{align}
C_{l}^{m}(a,b) & = e^{-i 2\pi \delta_{l}(k_m(f_a)
- f_a k_{m}^{'}(f_a))} C_{0}^m(a,b - \delta_{l} k_{m}^{'}(f_a)) \nonumber \\
& = e^{-i \delta_{l}\phi_m(f_a)} C_{0}^m(a,b - \delta_{l} k_{m}^{'}(f_a)) \label{eq:cwt_dispersion_para}
\end{align}
where $\delta_{l} = z_l - z_{0}$ denotes the spacing of the receiver $l$ from the reference $0$ and where $\phi_{m}(f_a)
\doteq 2\pi (k(f_a) - f_a k'(f_a)) = 2\pi f_a(s^{\phi}(f_a) - s^g(f_a)) $ is a phase factor. Thus for a given mode the CWT coefficients of
the mode at a given scale undergo a time shift according to the group slowness across the receivers along with a complex phase shift
proportional to the difference of the phase and the group slowness.  These ideas are depicted in Figure~\ref{fig:joint_schematic}(a) \& \ref{fig:joint_schematic}(b).   It becomes readily apparent that if the CWT coefficients corresponding to such a mode could be isolated and time shifted by a given moveout, then the phase slowness $s^{\phi}_m(f_a)$ can be extracted by solving for an exponential model when the moveout equals the group slowness $s^g_m(f_a)$.  

\begin{figure*}
\centering \makebox[0in]{
    \begin{tabular}{c}
      \includegraphics[height= 2.5 in]{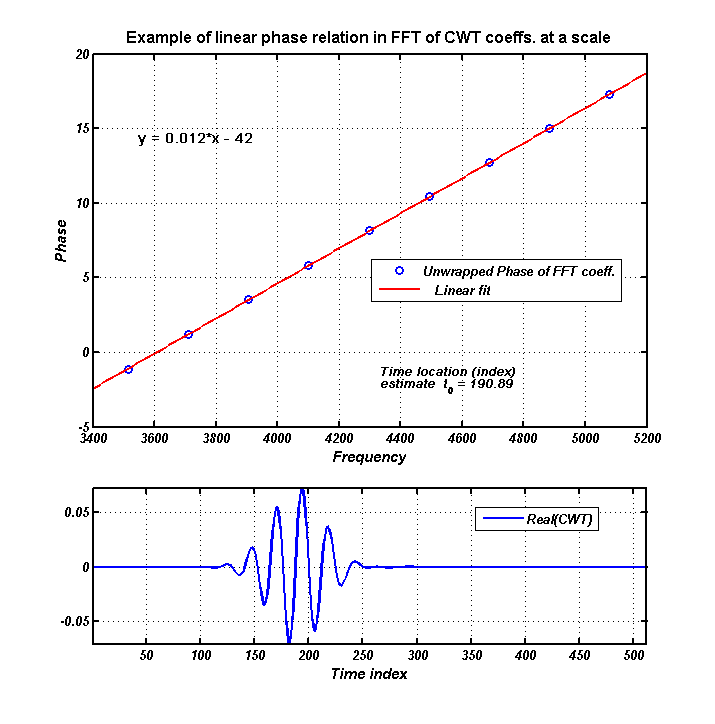}
      \end{tabular}}
  \caption{An example of the implication of time compactness of the CWT coefficients at a given scale at one receiver on the phase of the FFT coefficients in the band.  The lower panel depicts the real part of a typical CWT coefficient  of a time compact synthetic signal at a certain scale $a$ and plotted as a function of the shift parameter treated as a time index.  The upper panel shows the unwrapped phase of the FFT of the CWT coefficient taken with respect to the time index and plotted as a function of frequency.  This latter function manifests as a straight line with the slope related to the time location of the envelope peak of the CWT coefficients in the lower panel.  The  time location estimate shown (190.89)  is derived from the slope and is in good agreement with the true location of the envelope peak.  This approach is used to estimate the time location of time-compact modes in the received data by application to the amplitude coefficients $\bX-{F_a}$ obtained by solving  equation\ref{eq:OPT_fk}. }
  \label{fig:linear_phase}
\end{figure*}

The linearity of the CWT guarantees that in the presence of multiple modes and noise, the CWT of the data at scale $a$ is the superposition of the CWT of each mode, i.e.,
\begin{align}
\label{eq:zt_Taylor}
C_l(a,b) \approx \sum_{m=1}^{M} e^{-i \delta_{l,}\phi_m(f_a)} C_{0}^m(a,b - \delta_{l} k_{m}^{'}(f_a)) + C_l^w(a,b)
\end{align}
\begin{align}
\label{eq:fk_Taylor}
C_{l}(a,f) \approx \sum_{m=1}^{M} G^{*}(af) S_m(f) e^{- i2\pi (k(f_a) + k'(f_a)(f - f_a)) z_l} + W(a,f), \,\,\, f \in F_a
\end{align}
where $M$ is the number of modes,  $C_l^w(a,b)$ is the CWT of the noise, and $W(a,f)$ is the corresponding Fourier transform.   We observe that in the general case with time overlapped modes, we can no longer approach the  problem in terms of shifting the CWT coefficients and solving a sum of exponentials model.  However we can still represent the CWT coefficients in terms of broadband propagators using the above equation as detailed in the next section.   First we make the following observations regarding the CWT coefficients of  time compact propagating modes as  typically seen in borehole acoustic logging:
\begin{itemize}
\item[1.] Time compactness of mode $m$ implies time compactness of the CWT coefficients $C_{l}^{m}(a,b)$ at each scale. The time-frequency support of the modes in the CWT domain depends on the time-frequency uncertainty relation corresponding to the mother wavelet used - at higher frequencies (smaller scales) the time resolution is sharp but the frequency resolution is poor and vice-versa.
\item[2.] Time compactness in turn implies a linear phase relationship across frequencies $f \in F_a$ in the Fourier transform coefficients  $C_{l}^{m}(a,f)$ of the CWT data $C_{l}^{m}(a,b)$ at any receiver $l$. This is depicted in Figure~\ref{fig:linear_phase} for a synthetic example.
\end{itemize} 
It is worthwhile to note at this point that the first observation together with the dispersion relation of Equation \ref{eq:cwt_dispersion_para} in the CWT coefficients, formed the basis of EPRT processing as proposed in \cite{AeronICASSP08}, appropriate when there is a single dominant mode in the data or when the modes are well separated in time-frequency.

\subsection{Formulation in the CWT domain: sparse data representation using broadband propagators in $f-k$ and $z-t$ domains}
In this section we exploit the following aspects of mode characteristics, namely, (a) modes are nearly both time and frequency compact at scale $a$ and (b) the number of modes at scale $a$ is small. Then working under the assumption of the validity of the first order Taylor series approximation around the center frequency our methodology for broadband dispersion extraction is centered around the following \emph{constructive} representations of the CWT data at scale $a$ in $f-k$ and $z-t$ domains respectively.   

\subsubsection{Broadband propagator representation in $f-k$ domain over a frequency band $F_a$ at scale $a$}
In the following we will use a discrete set of frequencies $f_{1}, f_2,...,f_{N_{f}} \in F_a$ which correspond to the discrete DFT frequencies under the given sampling rate in the temporal domain. It is assumed that the temporal sampling is done at at-least the Nyquist rate. 
\begin{defn}
\label{def:broad_prop} A broadband propagator $\mathbf{P}_F(k(f_a),k'(f_a))$ (in band $F_a$) corresponding to a given phase slowness $\frac{k(f_a)}{f_a}$ and group slowness
$k'(f_a)$ is a block diagonal matrix $\in \Complex^{L\cdot N_f \times N_f}$ given by
\begin{align}
\mathbf{P}_F(k(f_a),k'(f_a)) &= \begin{bmatrix} \bphi(f_1)&  & & \\ & \bphi(f_2) & & \\ & & \ddots& \\ & & & \bphi(f_{N_f})
\end{bmatrix}
\end{align}
of sampled exponential vectors $\bphi(f)$ given by
\begin{align}
\mathbf{\bphi}(f) = \begin{bmatrix}
   e^{-i2\pi (k(f_a) + k'(f_a)(f - f_a) )(z_1 - z_0)} \\
  e^{-i2\pi (k(f_a) + k'(f_a)(f - f_a) )(z_2-z_0)} \\
  \vdots \\
  e^{-i2\pi (k(f_a)+ k'(f_a)(f - f_a) )(z_L - z_0)}
\end{bmatrix}
\end{align}
where $z_0$ is a reference receiver.
\end{defn}

From Equation~(\ref{eq:fk_Taylor}) one can easily see that the data in a given band is a \emph{superposition of $M$ broadband propagators} $\mathbf{P}_F(k_m(f_a),k_{m}^{'}(f_a)) $. As proposed in \cite{AeronTSP2011} using this broadband representation in the $f-k$ domain, a strategy to estimate the dispersion curves in the band $F_a$ consists of the following. 

\begin{itemize}
\item[1.] At scale $a$ form an \emph{over-complete dictionary} of broadband propagators $\mathbf{P}_F(k(f_a),k'(f_a))$ in
the $(f-k)$  domain spanning a range of group and phase slowness.

\item[2.] Assuming that the broadband signal is in the span of the broadband propagators from the over-complete dictionary, the presence of a few significant modes in the band implies that the signal representation in the over-complete basis is \emph{sparse}. In other words that the signal is composed of a superposition of few broadband propagators in the over-complete dictionary.

\item[3.] The problem of slowness dispersion extraction in the band can then be mapped to that of finding the sparsest signal representation in the over-complete
dictionary of broadband propagators.
\end{itemize}

The over-complete dictionary is constructed using a range of phase slownesses $s_{i}^{\phi}(f_a) = \frac{k_{i}(f_a)}{f_a}$ at the center frequency $f_a$ and a range of group slowness $k_{j}^{'}, j =1,2,...,n_2$ for the given band. The over-complete dictionary of broadband propagators for acoustic signal representation in a band can then be written as
\begin{align}
\bPhi_{F}^{a} = \begin{bmatrix} \Phi_1(F_a) \;|\; \Phi_2(F_a)\;|\; ... \;|\; \Phi_N(F_a)  \end{bmatrix} \in \Complex^{L.N_f \times N.N_f}
\end{align}
where $\Phi_{i + n_1\cdot(j-1)}(F_a) = \mathbf{P}_{F}(k_i(f_a),k_{j}^{'}(f_a))$ and $N = n_1 \times n_2$ is the number of broadband basis elements in the over-complete dictionary.

It was shown in \cite{AeronTSP2011} that assuming that the constructed dictionary contains elements corresponding to the phase and group slowness of the modes of interest, the CWT of the noisy data in band $F_a$ denoted by  $$\bY_{F_a} = [\bY_a(f_1) \bY_a(f_2) \hdots
\bY_a(f_{N_f})]^{T} \in \Complex^{L \cdot N_f}$$ at $L$ receivers can be written as
\begin{align}
\label{eq:lin_inverse} \bY_{F_a} = \bPhi_{F}^{a} \bX_{F_a}(:) + \mathbf{W}_{F_a},\,\, \bX_{F_a} \in \Complex^{N_f \times N}
\end{align}
for some unknown coefficient matrix $\bX_{F_a}$ whose $i + n_1\cdot(j-1)$-th column corresponds to the CWT coefficient vector at the reference receiver $z_0$, $ [C^{(i,j)} (a,f_1), C^{(i,j)} (a,f_2), ..., C^{(i,j)}(a,f_{N_f})]^ T$ corresponding to the mode with phase slowness $s_{i}^{\phi}$ and group slowness $s_{j}^{g}$. Here $(:)$ denotes the MATLAB $(:)$ operator which vectorizes the matrix by stacking columns on top of each other.  Since the number of modes $M << N$, $\bX_{F_a}$ is {\emph column sparse}, that is, only a few columns contain non-zero coefficient entries corresponding to the modes in the data.    It was shown in \cite{AeronTSP2011} that in that case, one can recover the columnn-sparse coefficient matrix $\bX_{F_a}$ and hence the modal dispersion by optimizing a sparsity penalized criterion, 
\begin{align}
\label{eq:OPT_fk}
\mbox{OPT\_fk}: \,\,\,
\hat{\bX}_{F_a} = \arg \min || \bY_{F_a} - \bPhi_{F_a} \bX_{F_a}(:)||_{2}^{2} + \lambda^{*} ||\bX_{F_a}||_{1,2}
\end{align}
where $|| \cdot ||_{1,2}$ denotes the $\ell_{1,2}$ norm, - computed by first forming the row-vector of $\ell_2$ norms of the columns of the matrix followed by taking the $\ell_1$ norm of the resulting row-vector - which has shown promise in recovery of sparse structured matrices, \cite{Tropp}. The optimal $\lambda^*$ can be chosen using the strategy presented in \cite{AeronTSP2011} or can be set to $ \sqrt{ M \log N \sigma^2}$ if an estimate of the number of modes $M$ and noise variance $\sigma^2$ per-dimension is available.

This approach to dispersion extraction was shown to be effective for recovering multiple modes overlapping in time-frequency (or time-scale) domain and produced robust and accurate estimates of the phase slowness at the center frequencies $f_a$ for each scale used for analysis.  However the group slowness estimates suffered from scatter as predicted by the Cramer Rao bounds for processing in the $f-k$ domain wherein we do not exploit any time domain compactness of the received signals.  In order to do that we turn to broadband propagators in the space-time domain which we now consider in this work.

\subsection{Space-time ($\lowercase{z-t}$) domain representation using time compact propagators with time-width $T_a$ at scale $a$ }
We define a  broadband propagator in space-time domain corresponding to a phase and group slowness as a time compact window propagating at the group slowness with a complex phase change across receivers in proportion to the difference of the phase and group slowness.  Mathematically, 
\begin{defn}
Let $\mathbf{u}_{T_a}(t)$ denote a rectangular window function of width $T_a$ centered at zero. Then the broadband
space time propagator at a scale $a$ is a real matrix of size $T \cdot L \times T$ and given by, 
\begin{align}
&\bU^{a}(s^{\phi},s^{g},t_0,T_a) = \begin{bmatrix}e^{i2\pi(s^{\phi} - s_g) f_a (z_1 - z_{0})}\mbox{diag}\left(\mathbf{u}_T(t - t_0 + s_g (z_1 - z_{0}))\right)\\
                                           \vdots\\
                                      e^{i2\pi(s^{\phi} - s^{g}) f_a (z_L - z_{0})}\mbox{diag}\left(\mathbf{u}_T(t - t_0 + s_g (z_L - z_{0}))\right)
                                     \end{bmatrix}
\end{align}
where diag($\cdot$) is the MATLAB operator and $z_0$ is a reference receiver.
\end{defn}

\begin{figure*}
\centering \makebox[0in]{
    \begin{tabular}{c}
      \includegraphics[height= 2 in]{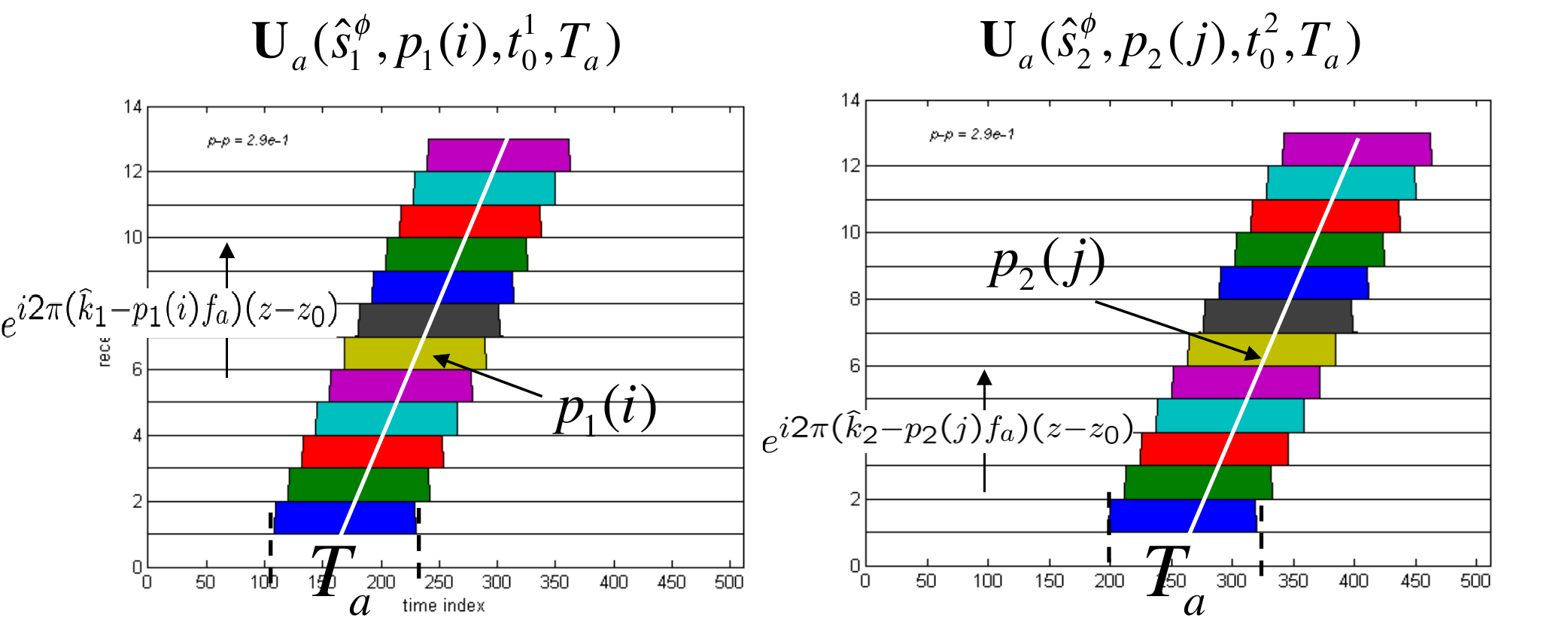}
      \end{tabular}}
  \caption{The broadband space time propagators in the continuous wavelet transform (CWT) domain.  }
  \label{fig:space_time_prop}
\end{figure*}

Examples of a broadband propagator in the space-time domain are shown in Figure~\ref{fig:space_time_prop}. From Equation~(\ref{eq:zt_Taylor}) it can be seen that the data in the space time domain can be written as a \emph{superposition of $M$ broadband propagators} $\bU^{a}(s_{m}^{\phi},s_{m}^{g},t_0^m,T_a)$. Then similar to the $f-k$ domain, using this broadband representation a similar strategy can be employed in the space-time domain. Namely, 

\begin{itemize}
\item[1.] At scale $a$ form an \emph{over-complete dictionary} of broadband propagators $\bU^{a}(s^{\phi},s^{g},t_0,T_a)$ in
the $(z-t)$  domain spanning a range of group and phase slowness \emph{and time locations}. 

\item[2.] Assuming that the broadband signal is in the span of the broadband basis elements from the over-complete dictionary, the presence of a few significant modes in the band implies that the signal representation in the over-complete dictionary is \emph{sparse}. In other words that the signal is composed of a superposition of few broadband propagators in the over-complete dictionary.

\item[3.] The problem of slowness dispersion extraction in the band can then be mapped to that of finding the sparsest signal representation in the over-complete
dictionary of broadband propagators.
\end{itemize}

The overcomplete dictionary is constructed using a range of phase slownesses $s_{i}^{\phi}(f_a) = \frac{k_{i}(f_a)}{f_a}$ at the center frequency $f_a$, a range of group slowness $k_{j}^{'}, j =1,2,...,n_2$ and \emph{in addition} a range of time points $t_{0}^{r}, r = 1,2,...,n_3$ indicating the possible locations of the modes. The over-complete dictionary of broadband propagators for acoustic signal representation in a band can then be written as
\begin{align}
{\cal U}_a = \begin{bmatrix} \Psi_1(T_a) \;|\; \Psi_2(T_a)\;|\; ... \;|\; \Psi_{N'}(T_a)  \end{bmatrix} \in \Real^{L \cdot T \times N' \cdot T}
\end{align}
where $\Psi_{i + n_1\cdot(j-1) + n_1\cdot n_2\cdot(r-1)}(T_a) = \bU^{a}(s_{i}^{\phi},s){j}^{g},t_{0}^{r},T_a)$ and $N' = n_1 \cdot n_2 \cdot n_3 $ is the number of broadband propagators in the over-complete dictionary.

Let the CWT at scale $a$ of the noisy data at $L$ receivers be denoted by $\bY_a \in \Complex^{L \times T}$. This data can be written as
\begin{align}
\label{eq:CWTlin_inverse} \bY_a (:) = {\cal U}_a \bX_{T_a}(:) + \mathbf{W}(:),\,\, \bX_{T_a} \in \Complex^{T \times N'}
\end{align}
for some unknown coefficient matrix $\bX_{T_a}$ whose $i + n_1\cdot(j-1)+ n_1\cdot n_2\cdot(r-1)$-th column corresponds to the CWT coefficient vector of length $T$ at the reference receiver $z_0$ and at time locations $t_{0}^{r}$, $$ [C^{i,j,r} (a,t_1), C^{i,j,r} (a,t_2), ..., C^{i,j,r}(a,t_{T})]^ T$$ corresponding to the mode with phase slowness $s_{i}^{\phi}$ and group slowness $s_{j}^{g}$. Here $(:)$ denotes the MATLAB $(:)$ operator which vectorizes the matrix by stacking columns on top of each other. Since the number of modes $M << N'$, similar to the $f-k$ domain strategy one can solve for the estimation of modal dispersion and mode spectrum (CWT) using the following optimization,
\begin{align}
\mbox{OPT\_zt}: \,\,\, 
\hat{\bX} _{T_a} = \arg \min || \bY_a(:) - {\cal U}_a \bX_{T_a}(:)||_{2}^{2} + \lambda^{*} ||\bX_{T_a}||_{1,2}
\end{align}
for some optimal value of $\lambda^*$. 

%
%


\section{ Joint $\lowercase{f-k}$ and $\lowercase{z-t}$ processing of CWT data at scale $\lowercase{a}$}
\label{sec:New_Approach}

Of the two approaches outlined above, clearly the approach using the broadband propagators in the $z-t$ domain exploits all aspects of the modal properties, namely both time-frequency compactness and sparsity in the number of modes. Nevertheless it is seen at once that due to the larger size of the optimization problem in the space time domain solving for OPT\_zt is computationally very intensive and may not be feasible. Besides it requires one to pick a good range for the arrival time of the modes to control the size of the dictionary. On the other hand OPT\_fk is quite tractable. Strictly speaking in the $f-k$ domain, one can also incorporate time compactness of modes by imposing a linear phase constraint in the broadband $(f-k)$ processing. However this would again impose significant additional computational requirements and is not feasible in general for practical applications. This is particularly true when the computation has to be done at wellsite  where computation speed and therefore efficiency
is critical.   

\begin{figure*}
\centering \makebox[0in]{
    \begin{tabular}{c}
      \includegraphics[height= 3 in]{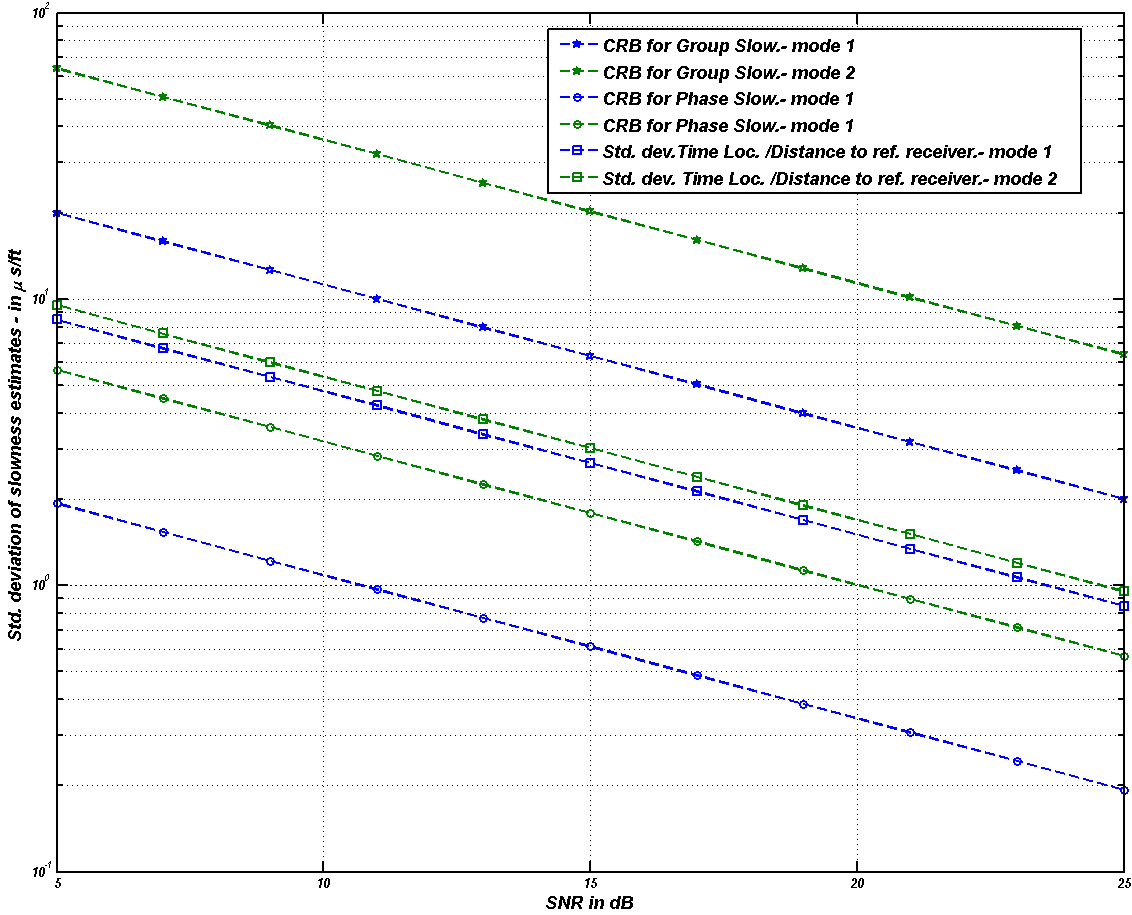}
      \end{tabular}}
  \caption{Figure showing the CRB for the slowness estimates for a two mode problem. One mode is 5-dB below the other mode. Note that the group slowness
estimates are less robust than the phase slowness estimates and time location estimates (normalized by the transmitter receiver spacing). }
  \label{fig:CRB_phaseslow_grpslow_timeloc}
\end{figure*}

Thus we propose a \emph{sequential} method for dispersion extraction in the CWT domain that utilizes the broadband multiple mode extraction methodology in the $(f-k)$ domain as proposed in \cite{AeronTSP2011} followed by the time compactness of modes in the  space-time domain in a manner similar to \cite{AeronICASSP08}. The main intuition is guided by a preliminary Cramer-Rao Bound (CRB) analysis carried out in Section~\ref{sec:CRB}, which shows that the broadband $f-k$ processing results in robust estimates of phase slowness and time locations of the mode while the group slowness estimates are not as robust, see Figure~\ref{fig:CRB_phaseslow_grpslow_timeloc} for an example on a synthetic data set. Here the variance in time location estimates of each mode are obtained using the CRB variance bounds on the corresponding coefficient phase estimates.  We note that it is possible to estimate the time locations by fitting a straight line to the coefficient phase across frequency as discussed below; the computed variance in the latter then yield the corresponding quantity for the former. 

Therefore in the sequential method below, we first estimate modal order, phase and group slownesses and the time locations of the modes using the OPT\_fk. Then we update the group slowness estimates by a simple exhaustive search over a range of group slownesses around estimated ones for each mode using the broadband space-time propagators.  The proposed methodology consists of 3 steps at each scale - 

\subsection{STEP 1: Simultaneous sparsity penalized broadband dispersion extraction in $f-k$ domain}
\label{sec:CWT_fk}
Solve OPT\_fk and pick an initial set of modes by picking say $R$ number of peaks from the vector $\mbox{colnorm}(\hat{\bX}_{F_a}) \in \Real^{1 \times N}$ and the corresponding phase and group slowness estimates along with the spectral estimates $\hat{C}_{z_0}^{r}(a,f)$ corresponding to the $r$-th column of $\hat{\bX}_{F_a}$. Proceed to the modal order selection and mode consolidation step below. 

\begin{remark}
The operation $\mbox{colnorm}(\cdot)$ for the matrix in the argument computes the $\ell_2$ norms along the columns and returns a row vector containing the norms. Mathematically for a matrix $\mathbf{M} \in \Complex^{m \times n}$ with entries $\mathbf{M}_{ij}$, $$\mbox{colnorm}(\mathbf{M}) = \left[ \sqrt{\sum_{j=1}^{m} |M_{1j}|^{2}}, \sqrt{\sum_{j=1}^{m} |M_{2j}|^{2}},..., \sqrt{\sum_{j=1}^{m} |M_{nj}|^{2}} \right]$$
The peak picking is done via an automatic procedure based on the relative amplitudes in vector $\mbox{colnorm}(\hat{\bX}_{F_a})$. 
\end{remark}

\subsection{STEP 2: Model order selection and mode consolidation}
In the broadband $(f-k)$  processing proposed in \cite{AeronTSP2011} we performed the model order selection based on clustering
in the phase and group slowness domain. In contrast here we will perform the model order selection and mode consolidation in the
phase slowness and time location domain. In order to obtain the time location estimates we use the fact that \textbf{time compactness
of the modes} imply a linear phase relationship across frequency in the mode spectrum. Based on the linear phase relationship
one can obtain estimates of the time location of the modes at scale ``a'' from estimates of the mode spectrum at the reference
receiver by fitting a straight line through the unwrapped phase of the estimated mode spectrum in the band. The slope of this
line is related to the index of the time location estimate of the mode via the relationship,
\begin{align}
\hat{t}_{0}^{r} = \dfrac{\mbox{slope(Unwrap. Phase $\hat{C}_{z_0}^{r}(a,f)$)}}{2\pi.T_s.10^{-6}}
\end{align}
where $T_s$ is the sampling time in $\mu s$.  This is illustrated in figure~\ref{fig:linear_phase}.

\begin{figure*}
\centering \makebox[0in]{
    \begin{tabular}{c}
      \includegraphics[height= 3.5 in]{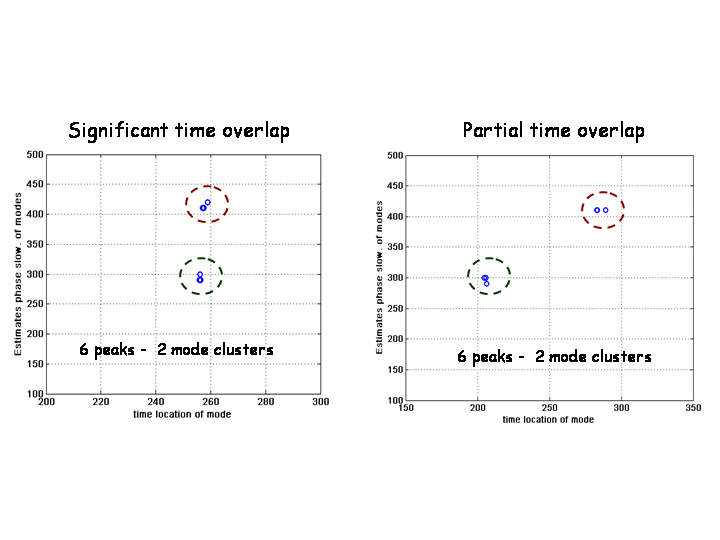}
      \end{tabular}}
  \caption{An example of mode consolidation and model order selection using the phase slowness and time location estimates. The phase slowness is measured in
$\mu s/ft$ and the time location is given in terms of sample number.  }
  \label{fig:mode_consolidation}
\end{figure*}

Example of  mode clustering in the phase slowness and time location domain for a synthetic two mode case for two scenarios of
partial time overlap and total time overlap are shown in Figure~\ref{fig:mode_consolidation}. The $R$ modes are clustered by a simple $k-$means clustering algorithm mode consolidation consists of averaging of the mode spectrum and the phase and group slowness estimates within each cluster. We denote the resulting model order by $M_a$. The time location estimates are not averaged but, for each consolidated mode the time location estimates are then estimated from the location of the \emph{peak} of the corresponding envelope in the time domain.

Steps 1. and 2. are summarized in Table \ref{tab:mode_clustering_phase_time}.

\begin{table}
\begin{center}
\begin{tabular}{|p{5.75in}|}
     \hline \\
     STEP 1. Pick a certain number of peaks say $R$ corresponding to the largest values in $\mbox{colnorm}(\hat{\bX}_{F_a})$ from the estimated solution from $f-k$ processing. \\ \\
     Step 2. Estimate time locations of the \emph{modes} corresponding to each of these peaks by fitting a straight line through the phase of the
     mode spectrum coefficients.  \\ \\
     STEP 3. Perform k-means clustering in the phase slowness and time location domain. \\ \\
     STEP 4. Declare the resulting number of clusters, $M_a$ as the model order (at scale $a$). \\ \\
     STEP 5. Mode consolidation - \\
     \hspace{5mm} 5a. The mode spectrum corresponding to each cluster is obtained by summing up the estimated mode spectrum coefficients corresponding to the
points in the cluster. \\
     \hspace{5mm} 5b. The slowness dispersion estimates are taken to be the average over the slowness dispersion parameters corresponding to the cluster points. \\
     \hspace{5mm} 5c. For each consolidated mode the time location estimates are then to be corresponding to the \emph{peak} of the envelope in the time domain.\\\\
     \hline
\end{tabular}
\caption{Table illustrating the steps for model order selection and mode consolidation in the phase slowness and time location
domain. }
\label{tab:mode_clustering_phase_time}
\end{center}
\end{table}

\subsection{STEP 3: Space-time processing of the consolidated modes for refining mode spectrum and group slowness estimates}

\begin{figure*}
\centering \makebox[0in]{
    \begin{tabular}{c}
      \includegraphics[height= 2.75in]{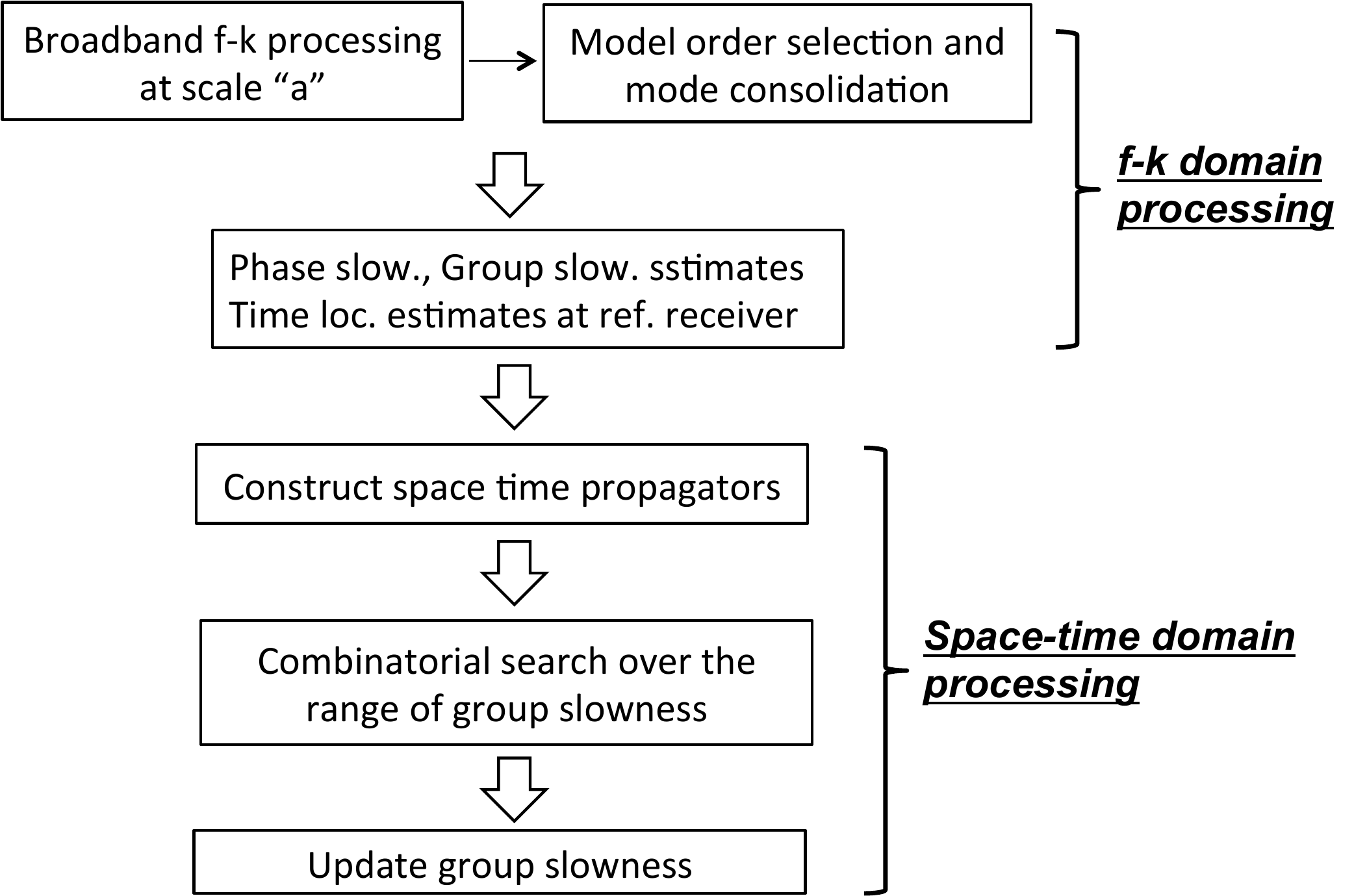}
      \end{tabular}}
  \caption{The flow of the processing in the space time domain processing the output of the broadband processing
  in the $(f-k)$ domain as applied to the CWT coefficients at a given scale. }
  \label{fig:flow_fk_tz}
\end{figure*}
Let the declared number of modes from $(f-k)$ processing at scale $a$ be $M_{a}$. Let the corresponding phase slowness estimates
be given by $\hat{s}_{m}^{\phi}$ and let the time location estimates be given by $\hat{t}_{0}^{m}$ for $m = 1, 2,...,M_{a}$. The
time window width $T$ for each mode around the time location $\hat{t}_{0}^{m}$  is based on the effective time width of the
analyzing wavelet at scale $a$. We pick a range of test moveouts for each mode around the estimated group slowness from the $(f-k)$ processing. 
For an $M_a$-tuple of test moveouts say $\mathbf{p} = [p_1, p_2, ..., p_{M_a}]$ we form space time propagators using the phase slowness and time location estimates obtained from $(f-k)$ processing.
Then given the CWT array data $\bY_a$ at scale $a$ we form the following system of equations.
\begin{align}
\bY_a(:) = {\cal U}_a(\mathbf{p}) \bX_{T_a}(:) + \mathbf{W}_a(:)
\end{align}
In the above expression,  the matrix
\begin{align}
{\cal U}_a(\mathbf{p}) = [ \bU^{a}(\hat{s}_{1}^{\phi},p_{1},t_{0}^{1},T_a), ...,\bU^{a}(\hat{s}_{M_a}^{\phi},p_{M_a},t_{0}^{M_a},T_a)]
\end{align}
is the matrix of broadband propagators corresponding to an $M_a$-tuple of test move-outs corresponding to the $M_a$ modes. Note that here we have significantly reduced the problem size for the $z-t$ domain processing and we only refine over the group slowness estimates.  Note that this is an over-determined
system of equations and in order to estimate the CWT coefficients at the test moveout vector $\mathbf{p}$ we simply form the minimum mean squared error (MMSE) estimate under the
observation model. To this end for each $M_a$-tuple $\mathbf{p}$s define
\begin{align}
e(p_1,..,p_{M_a}) & = || \bY_a - {\cal U}_a(\mathbf{p}) \hat{\bX}_{T_a,\mathbf{p}}(:)||\\
 \hat{\bX}_{T_a,\mathbf{p}} (:)& =  {\cal U}_a(\mathbf{p})^{\#} \bY_a
\end{align}
where $e$ is the residual error for the $M_a$-tuple test moveouts and $(\cdot)^{\#}$ denotes the pseudo-inverse operation. For updating
the group slowness estimates we do a combinatorial search over all possible choices $M_a$ tuples of test moveouts, as dictated by the range chosen for each mode,  and pick the combination that minimizes the residual error, i.e.,
\begin{align}
[\hat{p}_1,...,\hat{p}_{M_a}] = {\arg\min}_{[p_1,..,p_{M_a}]} || \bY_a(:) - {\cal U}_a(\mathbf{p}) \hat{\bX}_{T_a,\mathbf{p}}||
\end{align}

\subsection{Overall processing flow for dispersion extraction}

Choose disjoint or partially overlapping frequency bands with given center frequencies and bands around the center frequencies.
Note that there are two choices for this depending on the whether one uses CWT of the data for dispersion extraction or not. With the CWT of the data the choice of the frequency bands corresponds to the center frequencies of the dyadic scales and the
effective bandwidth at those scales.

For each of the band do the following.

\begin{itemize}
\item[1.] Execute a broadband processing in band F -
\begin{itemize}
\item[1a.] Construct over-complete dictionary of broadband propagators $\bPhi_{a}$ in band $F_a$ for a given range of phase and group slowness and pose the problem
as the problem of finding sparse signal representation in an over-complete basis. 
\item[1b.] Solve the resulting optimization problem using the algorithm outlined in \cite{AeronTSP2011}.
\end{itemize}
\item[2.] Model order selection and mode consolidation - Execute steps in Table 1.
\item[3.] For each mode choose a range of move-outs around the group slowness estimates and \emph{build space-time propagators} using the phase slowness
estimates and  time location estimates of the modes.
\item[4.] For each $M_a$ tuple of move-outs, $\mathbf{p} = [p_1,...,p_{M_a}] $  find estimates $\hat{\bX}_{T_a,\mathbf{p}}$ of  CWT coefficients using 
$\hat{\bX}_{T_a,\mathbf{p}}(:) =  {\cal U}_a(\mathbf{p})^{\#} \bY_a$.
\item[5.] Update group slowness estimates -
\begin{align}
\hat{p}_1,...,\hat{p}_{M_a} = arg\min_{p_1,..,p_{M_a}} || \bY_a(:) - \bU_a(\mathbf{p}) \hat{\bX}_{T_a,\mathbf{p}}(:))||
\end{align}
and proceed to the next band.
\end{itemize}


\section{Performance on real data}
\label{sec:Real_Data}

We now test the performance of the proposed method on a representative set of real data examples drawn from a variety of scenarios using borehole sonic tools in both wireline and LWD conveyances.   In each of this cases we have at least two borehole mode arrivals overlapping in time.  Besides potentially presenting an issue for model based inversion that pre-supposes the existence of only a single mode in the processing window, the second mode might represent an opportunity for enhanced interpretation.  As explained before the relatively small number of receivers (12-13) and short array aperture of a few feet make it challenging to separate such closely spaced overlapped modes.

\begin{figure}
\centering \makebox[0in]{
    \begin{tabular}{cc}
    \includegraphics[height= 2in]{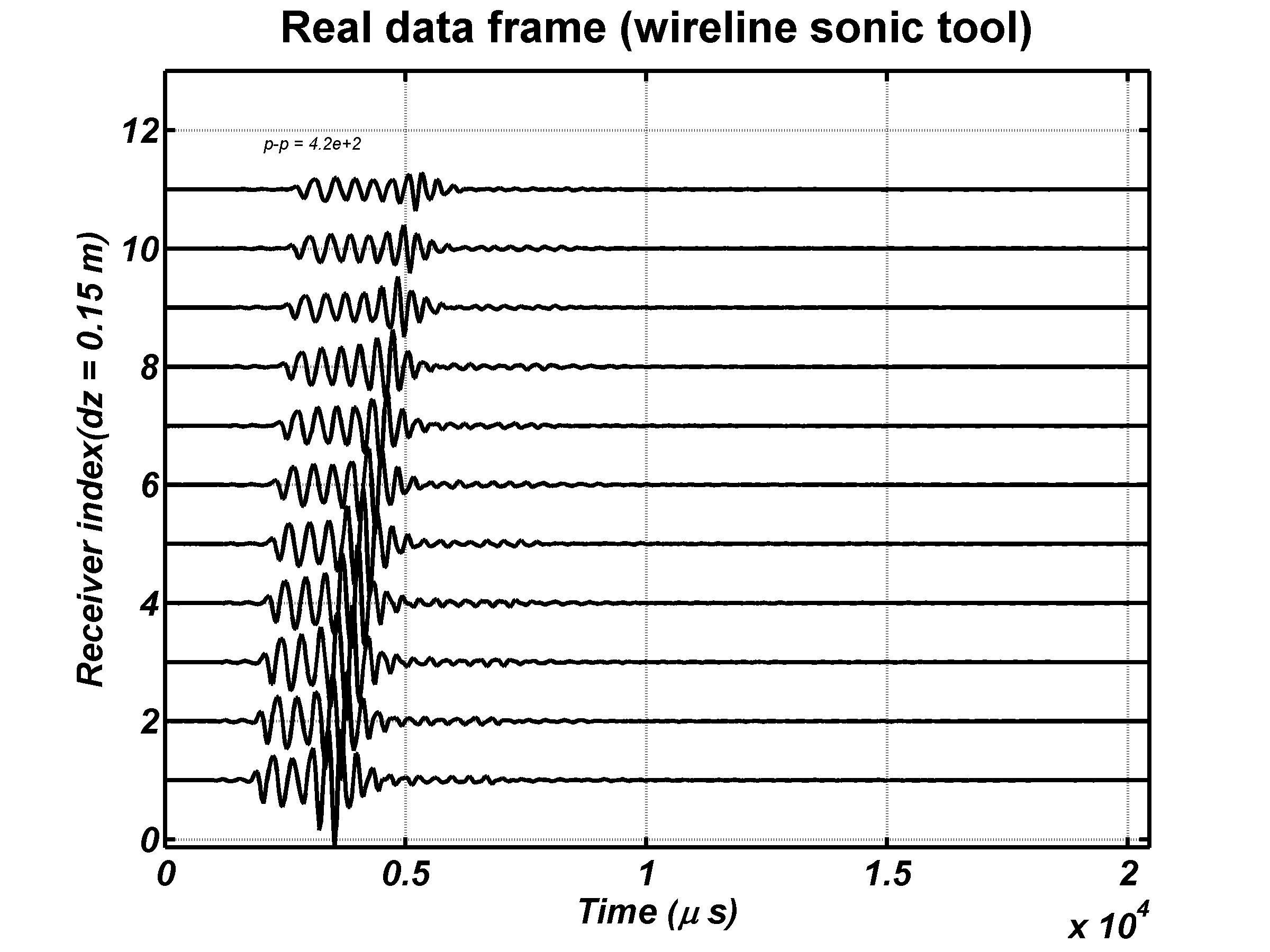} &
    \includegraphics[height= 2in]{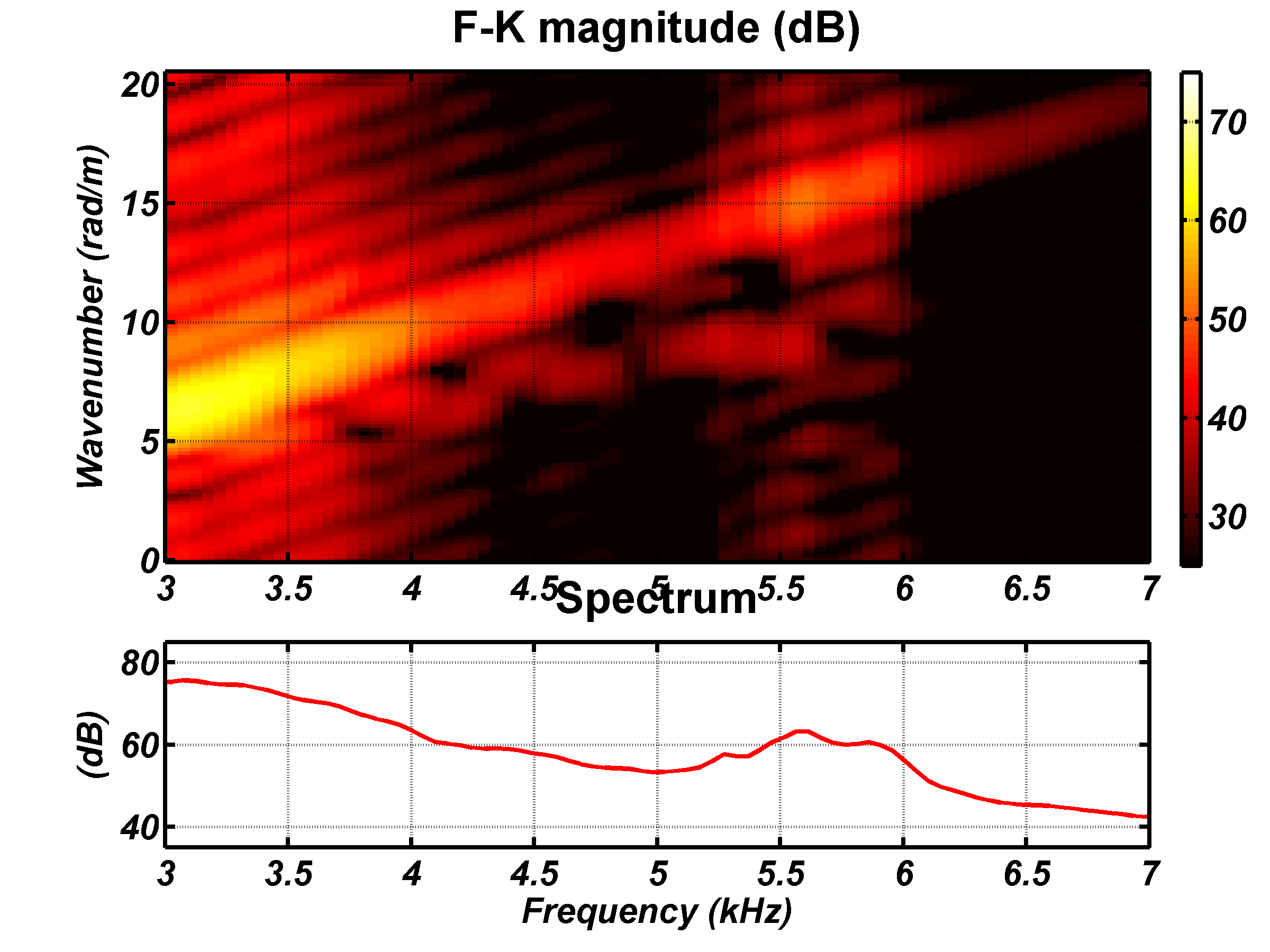} \\
(a) & (b) 
   \end{tabular}}
  \caption{An example of data  obtained with  a wireline sonic dipole acquisition showing plots of (a) traces of an array of waveforms and (b) corresponding  f-k and spectrum.  Note that we observe a single wavetrain which nevertheless suggests the presence of more than one mode which partially overlap in time and frequency. This is observed atlbeit weakly on the f-k plot.}
\label{fig:dipss_wave}
\end{figure}

\begin{figure}
\centering \makebox[0in]{
    \begin{tabular}{cc}
      \includegraphics[height= 2in]{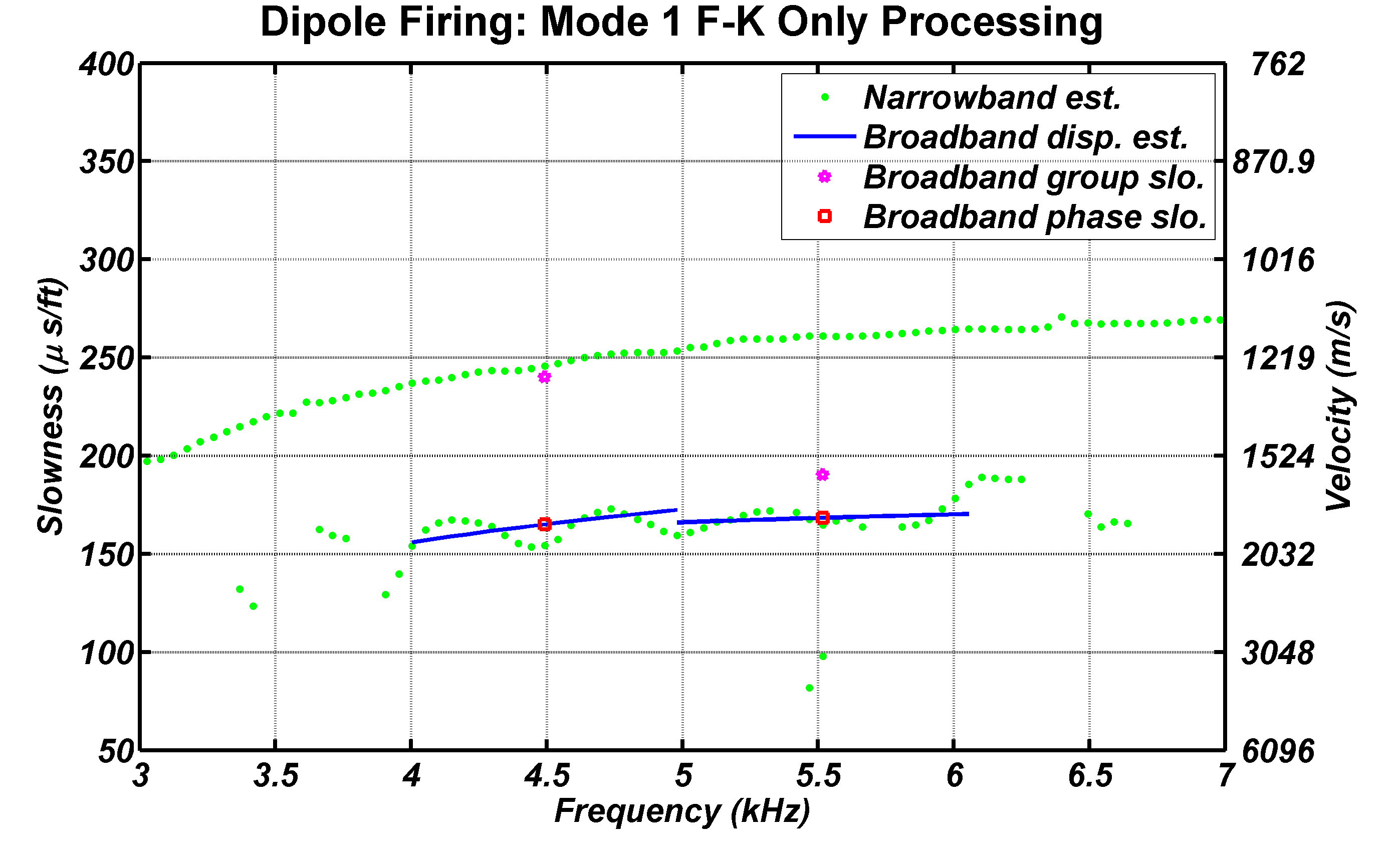} & \includegraphics[height= 2in]{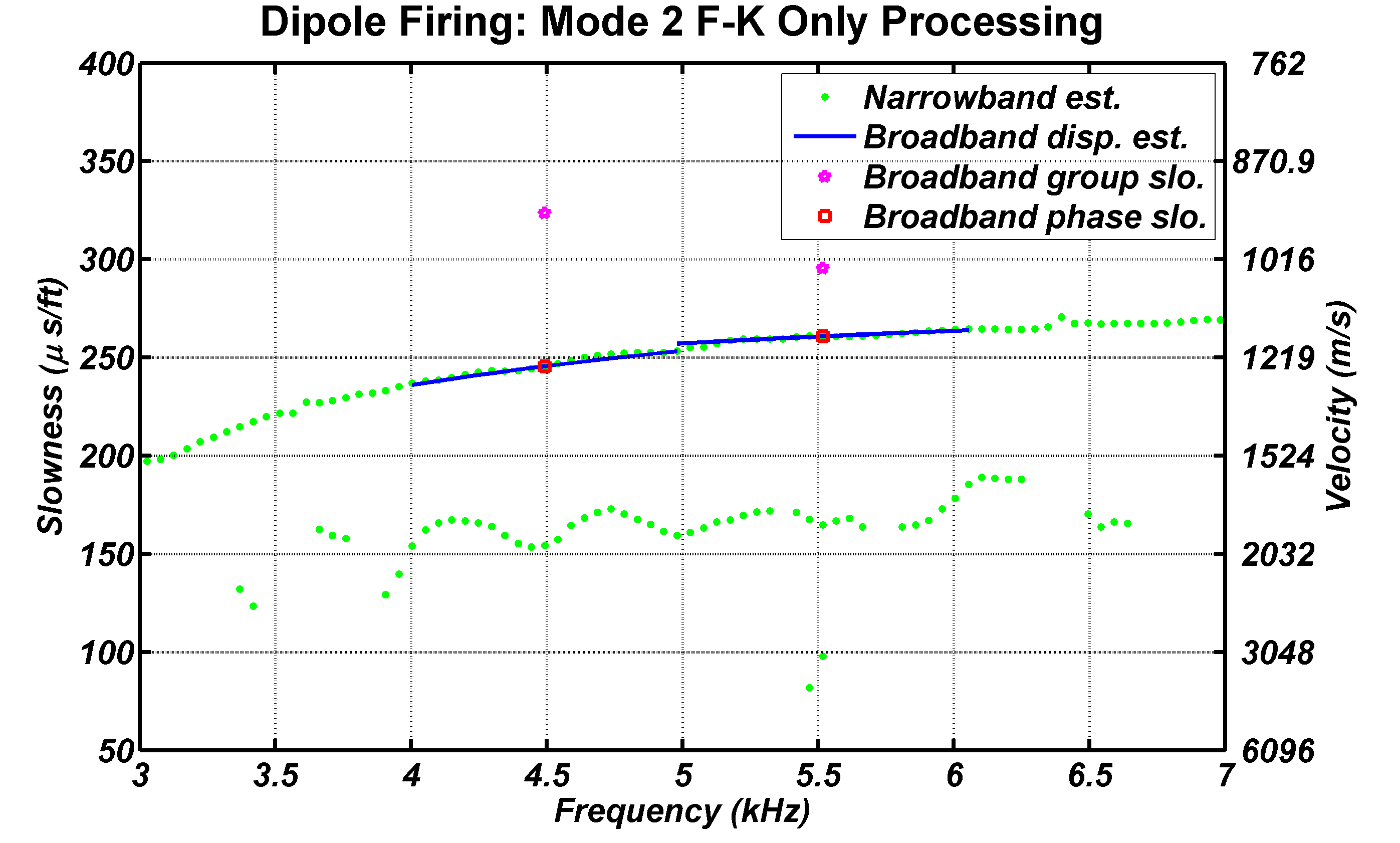}\\
      (a) &(b)\\
            \includegraphics[height= 2in]{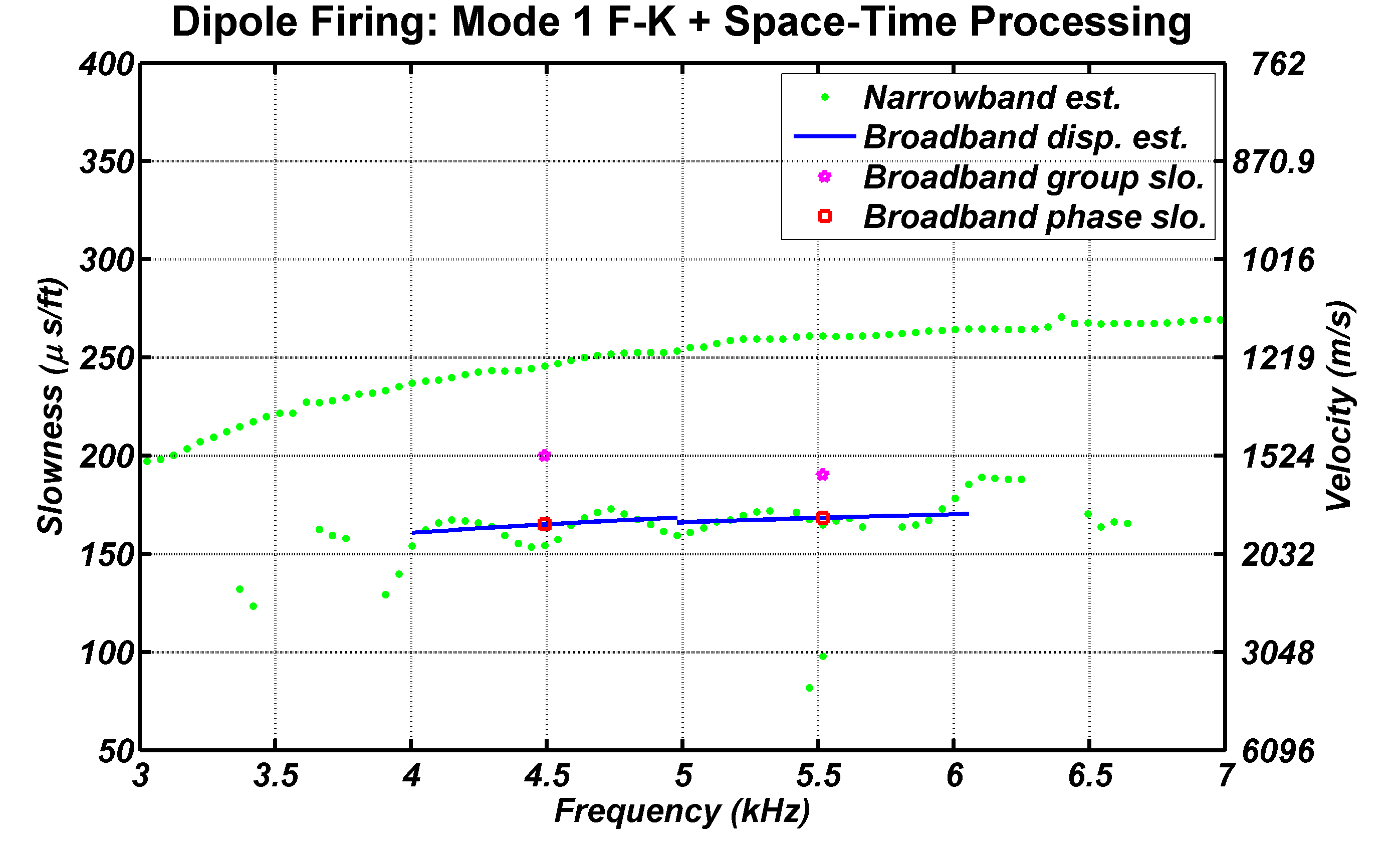} & \includegraphics[height= 2in]{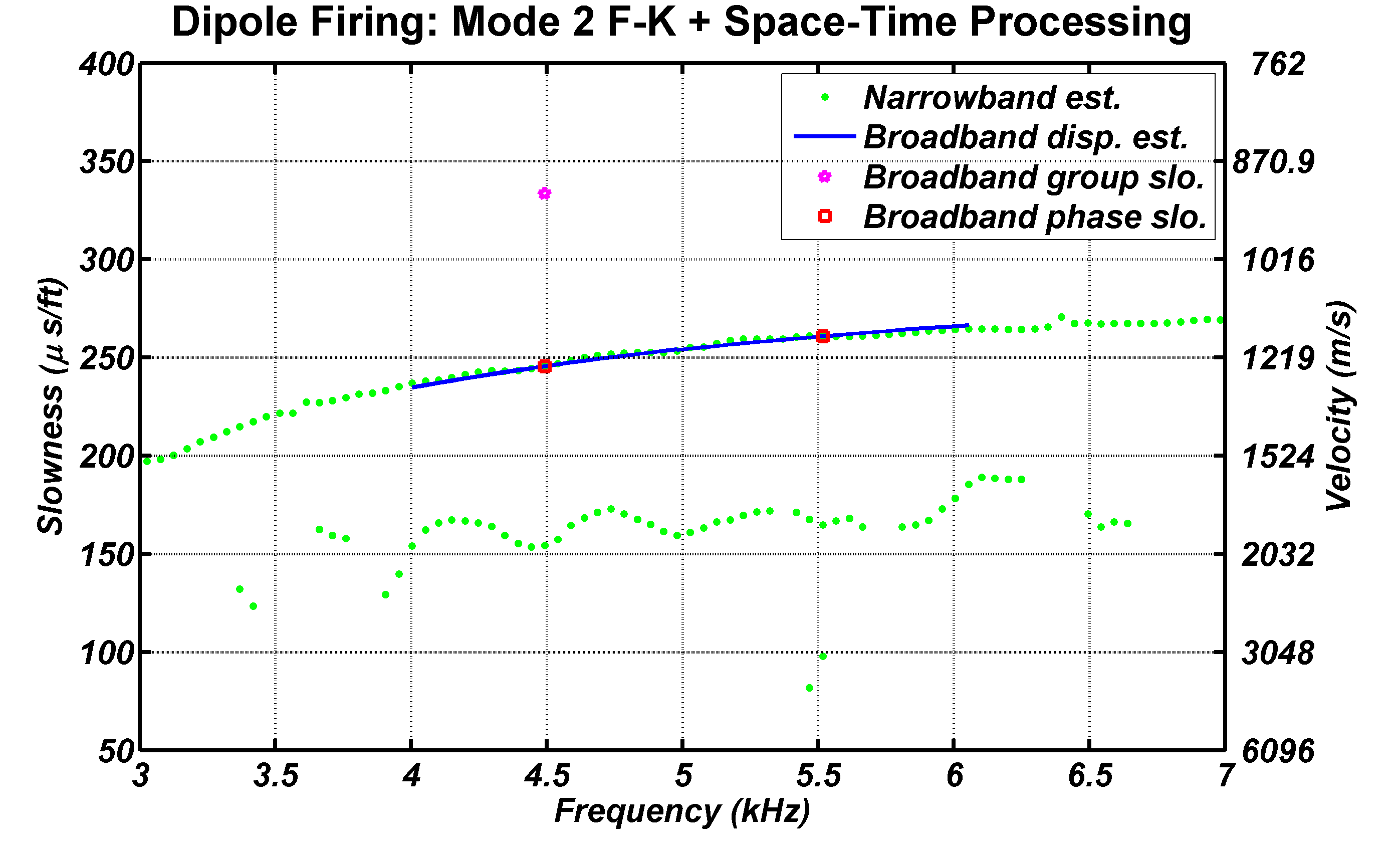}\\
(c)&(d)
      \end{tabular}}
  \caption{Dispersion extraction results for the wireline sonic dipole data shown in figure~\ref{fig:dipss_wave}. In each panel, the results of the narrowband processing are shown with green dots, indicating the presence of two modes in the data.  Note the scatter especially for the weaker higher order mode around $150 \mu s/ft$.  The broadband processing results using the $f-k$ processing only are shown in panels (a) and (b) for modes labelled 1 and 2 respectively. We observe that while the phase slowness estimates indicated by the squares at the CWT center frequencies appear to be accurate and consistent, there is somewhat greater scatter of the group slowness estimates indicated by the stars.  This leads to errors in the slope of the dispersion curves resulting in dispersion estimate errors at the CWT band edges and discontinuities in the extracted dispersions.  The results of the processing with the refinement of the space time processing are shown in panels (c) and (d) respectively.  The group slowness estimates are now improved resulting in much improved dispersion curve estimates without the discontinuous artifacts.}
  \label{fig:dipss}
\end{figure}

We first look at a case of a dipole acquisition with a wireline sonic tool as described in \cite{SonicScan06}.  Although the dipole firing typically produces a dominant lowest order flexural mode, sometimes additional modes such as higher order flexural modes are also excited.  These
may be partially  time overlapped with the principal mode and it is of interest in that case to be able to extract the overlapping dispersion curves for proper analysis.  Figure~\ref{fig:dipss_wave} shows an array of traces as well as the frequency-wavenumber ($f-k$) and spectrum display  for such a case of wireline dipole sonic data.  There appears to be more than one propagating modes in a single wavetrain.  This is confirmed in figure~\ref{fig:dipss} where we show the results of the dispersion extraction.  We observe that while the broadband approach results in significantly more stable estimates of the dispersion, the $f-k$ only processing induces discontinuous artifacts in the extracted dispersion due to errors in the group slowness estimates. The use of the space time processing to refine the group slowness estimates results in significant improvement in the quality of the dispersion curves extracted.

Next we turn to examples from logging while drilling (LWD) acquisitions which is of increasing importance for formation evaluation while inducing greater complexity in the acoustics.  For example, for the case of the dipole firing, the drill collar on which the tool is mounted plays a dominant role in the acoustic response.  In particular with a fast formation, both formation open hole flexural and drill collar tool flexural are strongly coupled  and produce two hybrid modes which moreover arrive in a time overlapped fashion\cite{Sinha09}.  Proper interpretation therefore requires extraction of both coupled mode dispersions. The problem is made more challenging by the fact that the mode dominated by the tool at low frequencies is much stronger making it harder to properly extract the faster weaker mode that is more sensitive to the formation of interest.

\begin{figure}
\centering \makebox[0in]{
    \begin{tabular}{cc}
    \includegraphics[height= 2in]{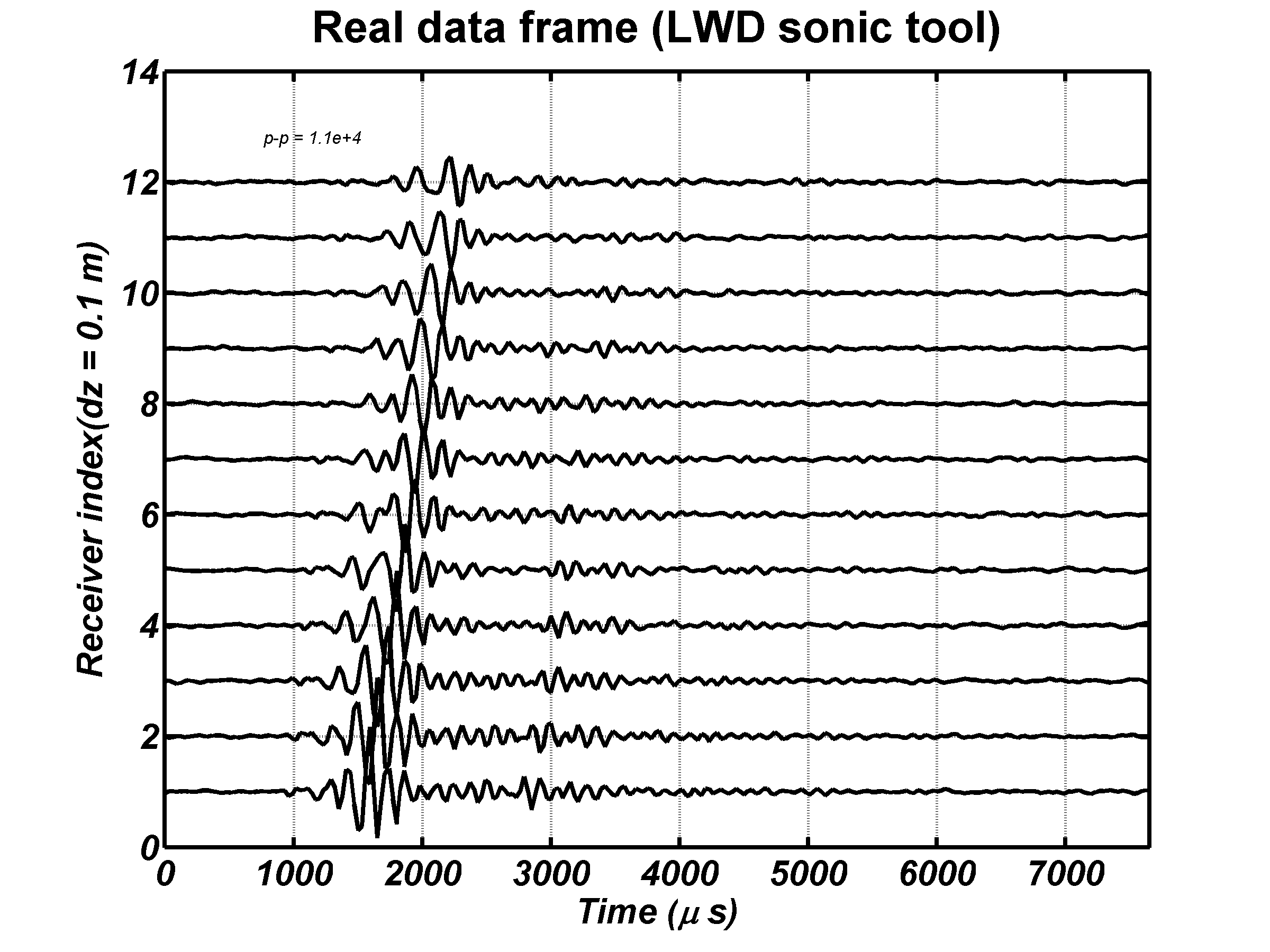} &
    \includegraphics[height= 2in]{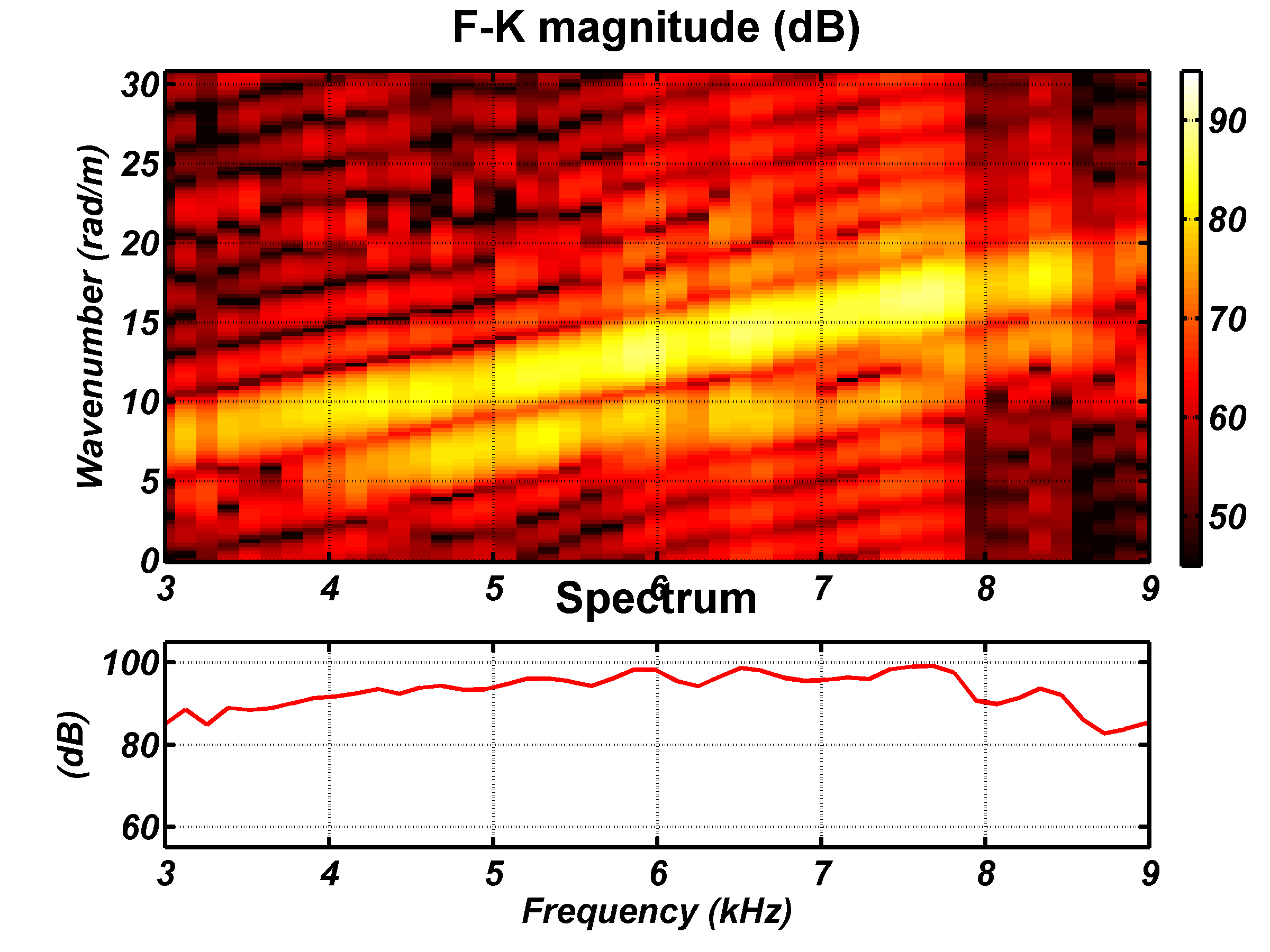}\\
  (a) & (b) 
   \end{tabular}}
  \caption{An example of data (Frame 1)  obtained with  a LWD  sonic dipole acquisition showing (a) traces of a waveform array of traces and (b) corresponding  f-k and spectrum on the right.  Note that we observe a single wavetrain comprising the two coupled modes.  The weaker later arrival is propagating at the same slowness and arises due to a reflection below the transmitter.}
\label{fig:Dipole1wave}
\end{figure}
\begin{figure}
\centering \makebox[0in]{
    \begin{tabular}{cc}
      \includegraphics[height= 2in]{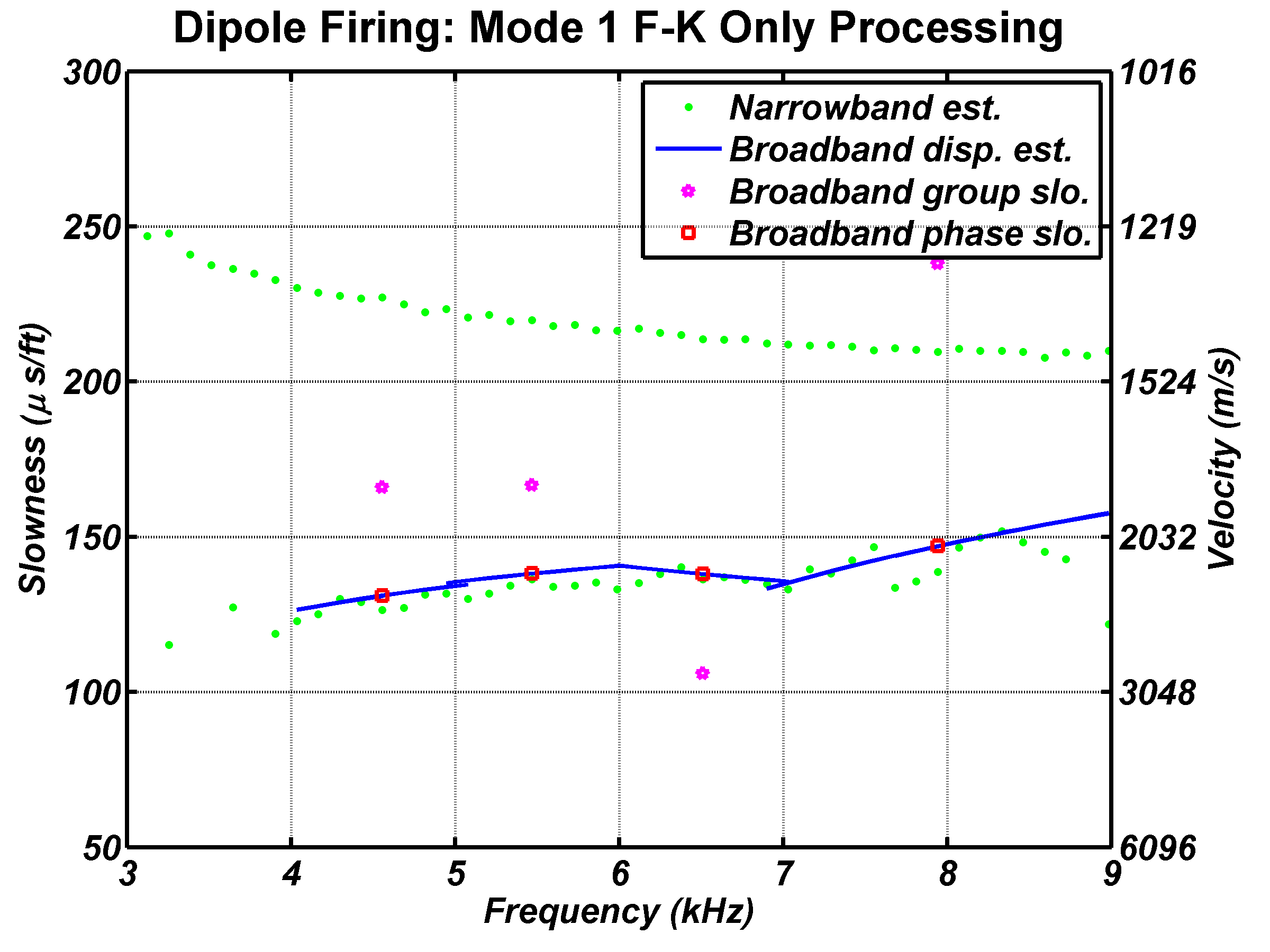} & \includegraphics[height= 2in]{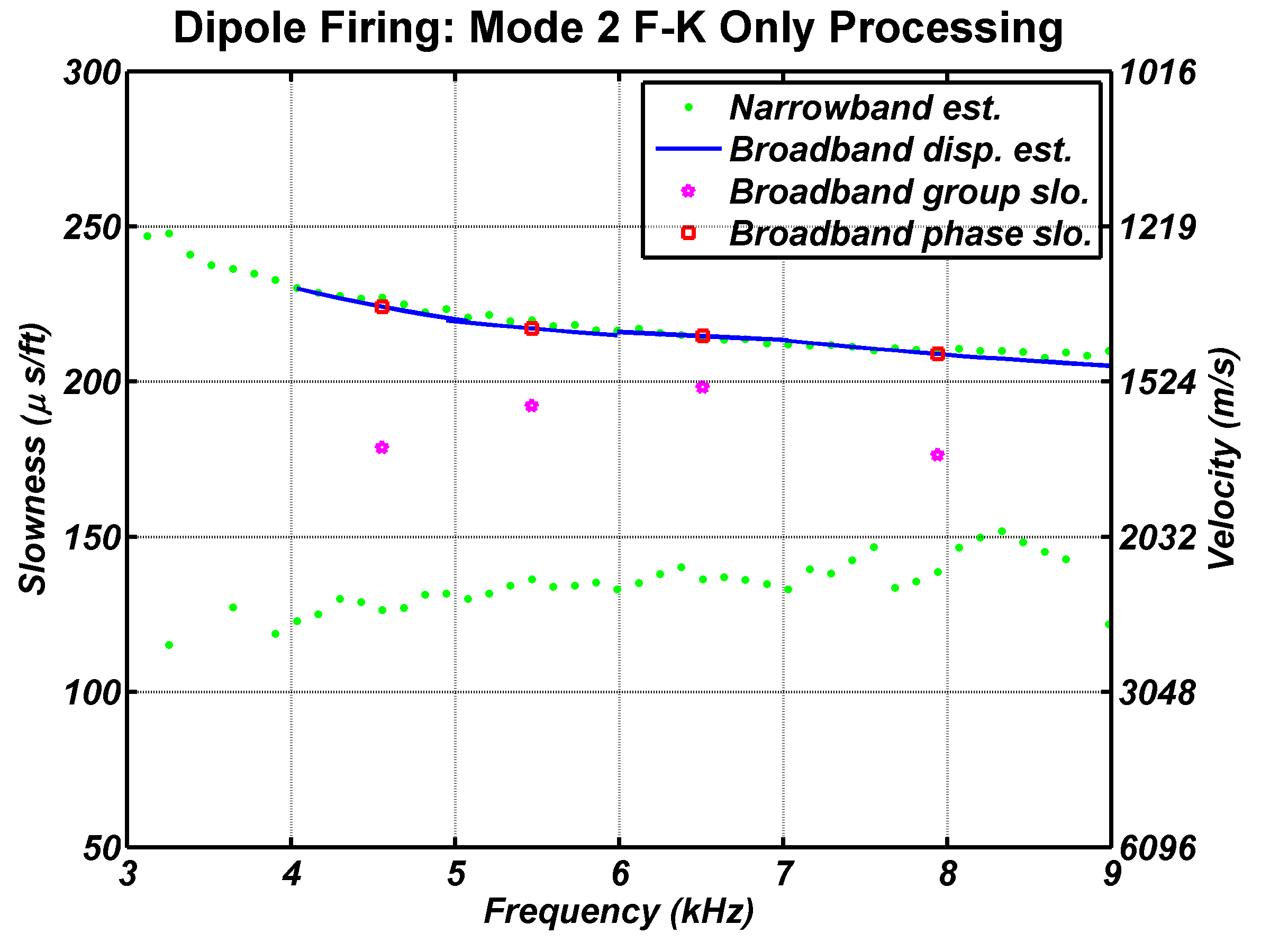}\\
      (a) &(b)\\
            \includegraphics[height= 2in]{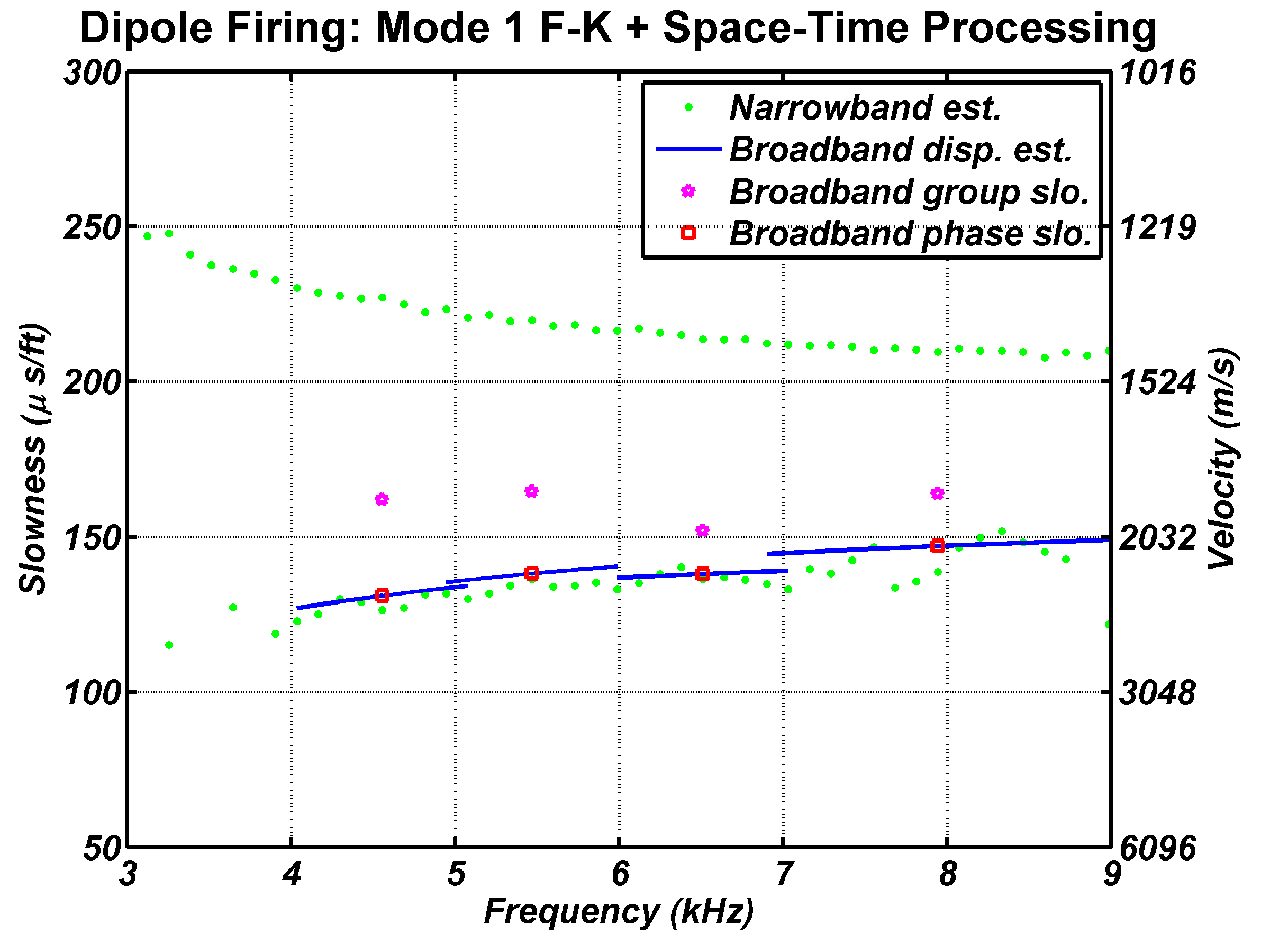} & \includegraphics[height= 2in]{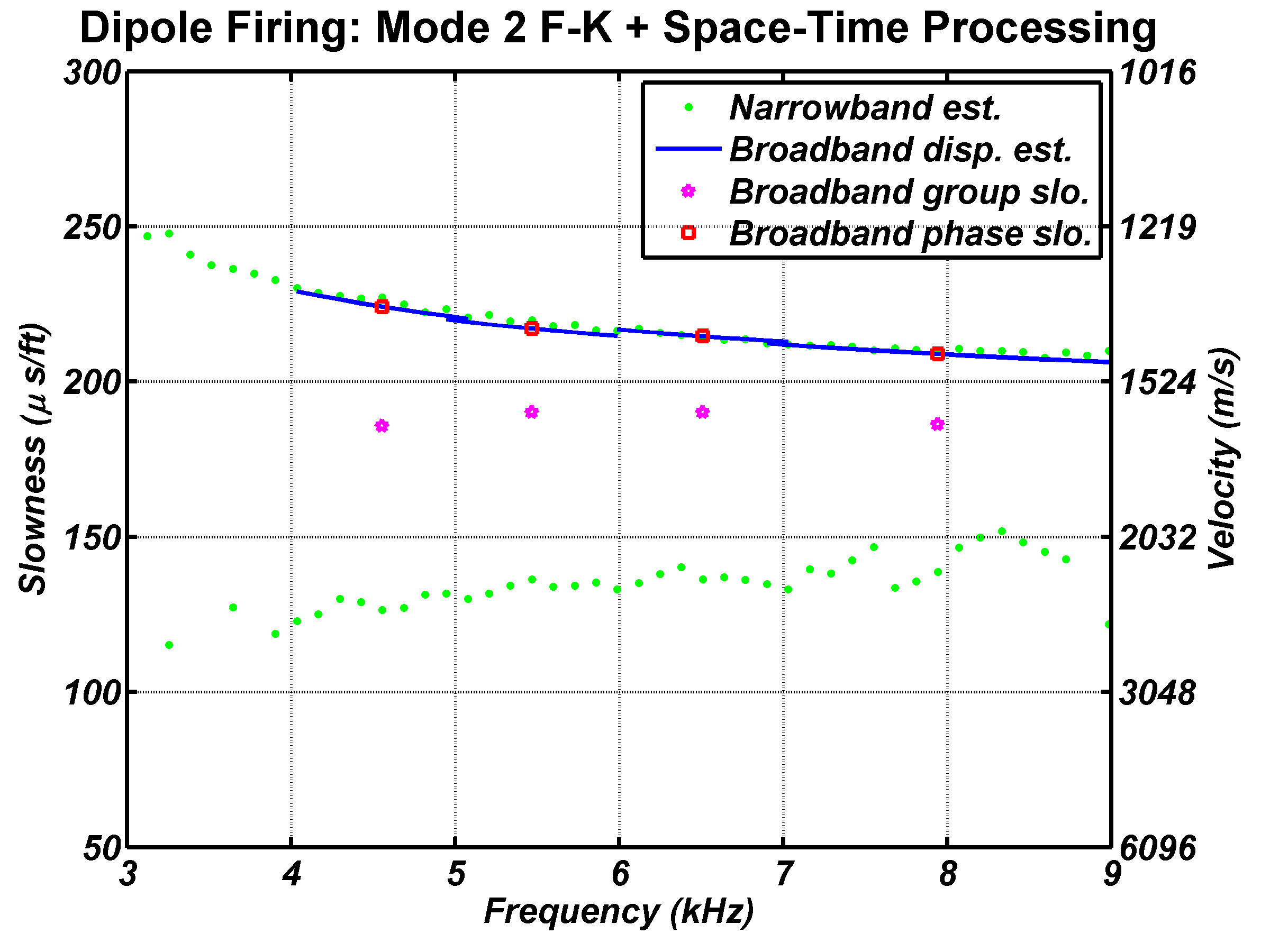}\\
(c)&(d)
      \end{tabular}}
  \caption{Dispersion extraction results for a LWD dipole firing - Frame 1. The annotations and conclusions are similar to those in figure~\ref{fig:dipss}.  The space-time processing clearly improves the group slowness estimates of the weaker (faster) mode.}
  \label{fig:Dipole1}
\end{figure}
\begin{figure}
\centering \makebox[0in]{
    \begin{tabular}{cc}
    \includegraphics[height= 2in]{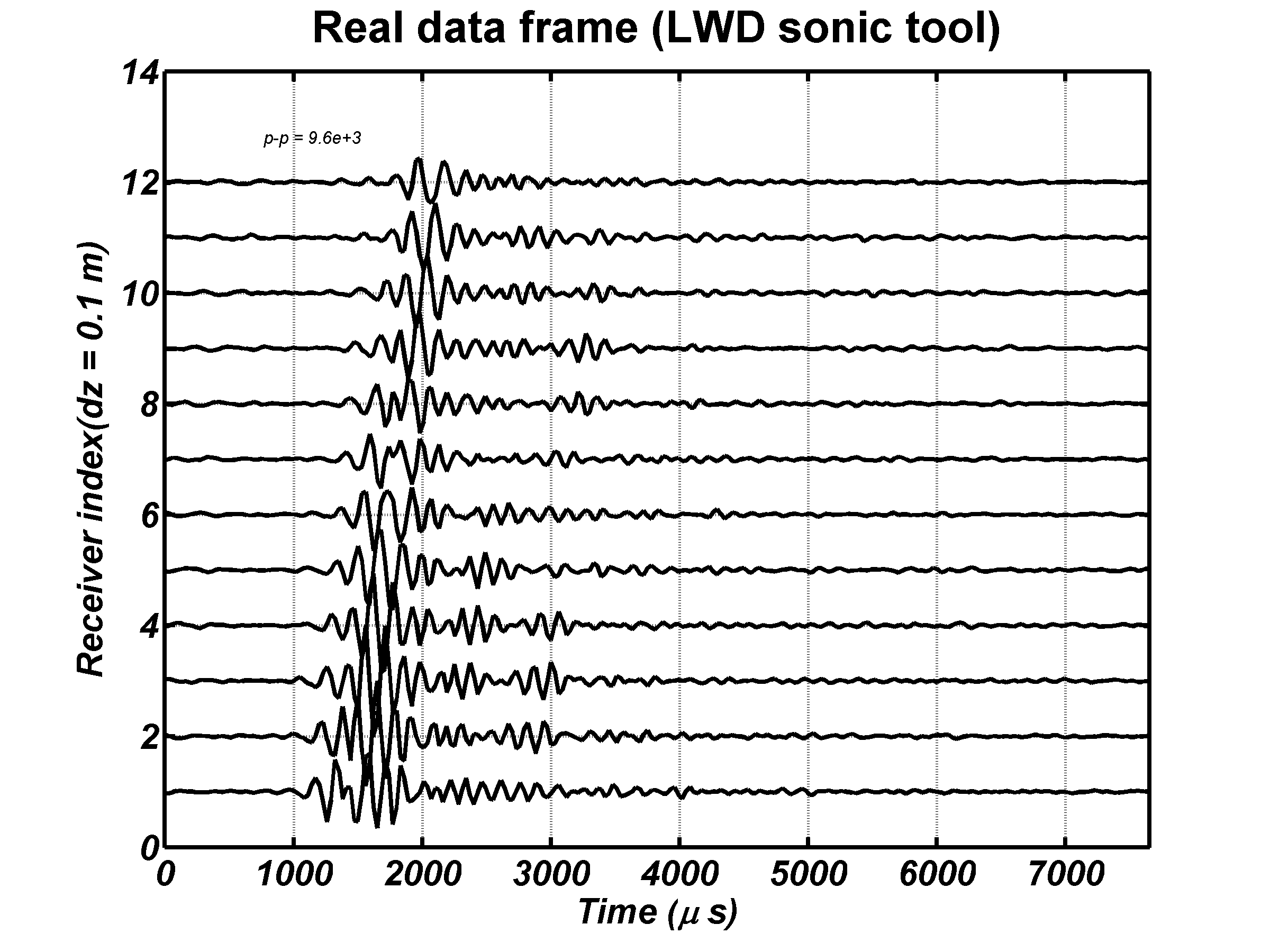} &
    \includegraphics[height= 2in]{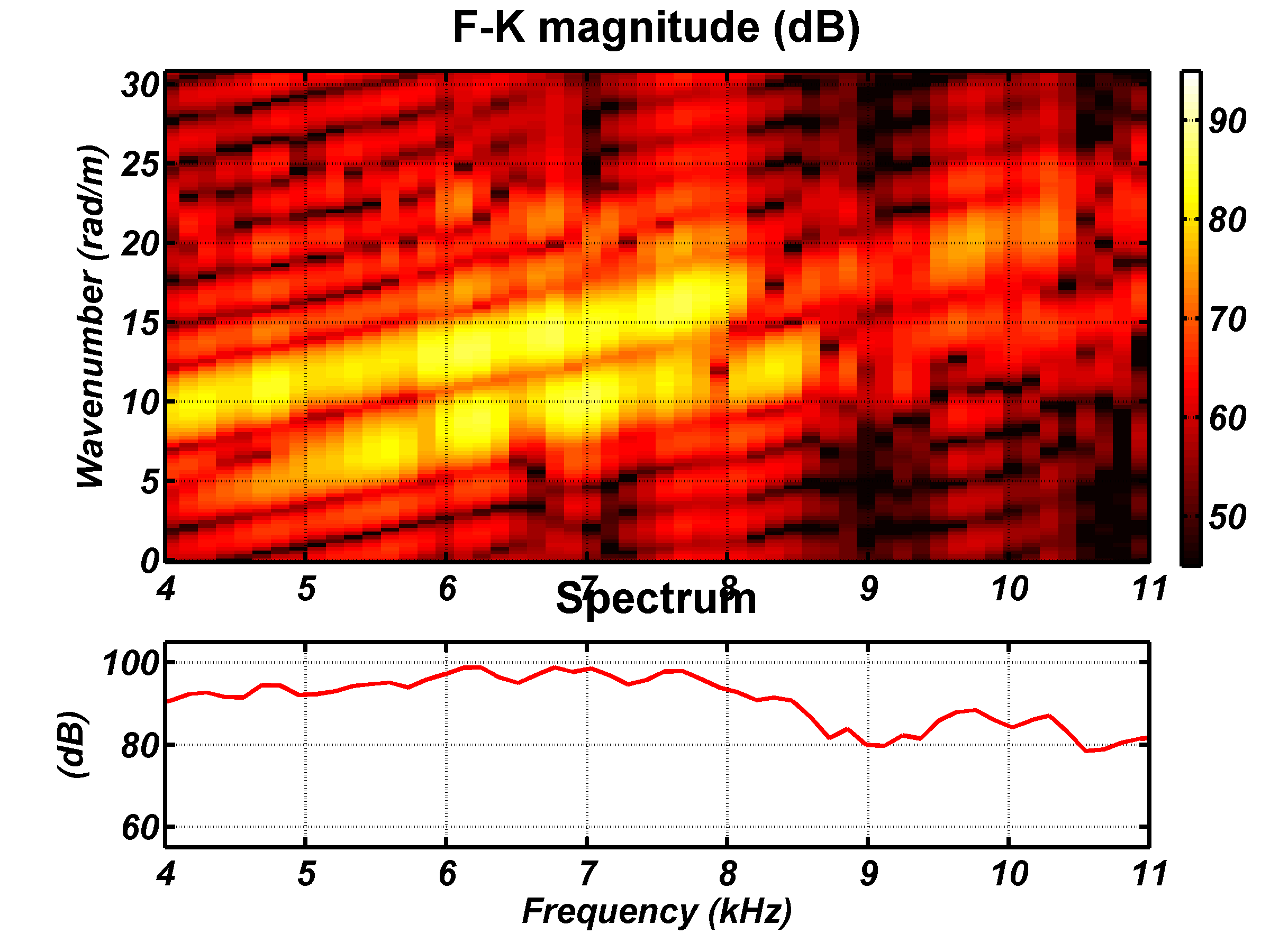}\\
  (a) & (b) 
   \end{tabular}}
  \caption{Another example of data (Frame 2)  obtained with  a LWD  sonic dipole acquisition showing plots of  (a) waveform array of traces on the left and (b)  f-k and spectrum on the right.  Here we observe that in addition to features in the previous example we see interference resulting in some incoherence across the array. }
\label{fig:Dipole2wave}
\end{figure}
\begin{figure}
\centering \makebox[0in]{
    \begin{tabular}{cc}
      \includegraphics[height= 2in]{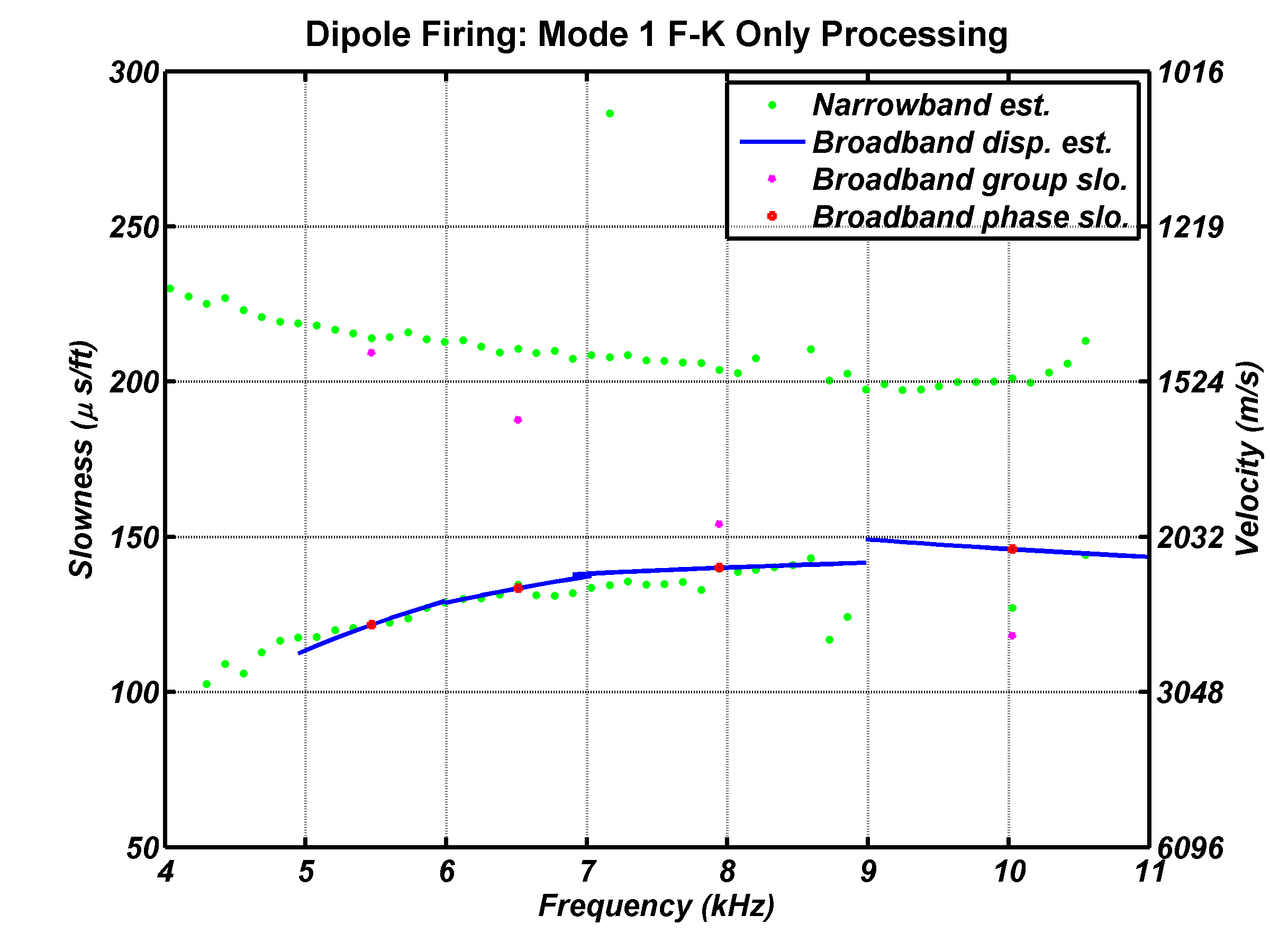} & \includegraphics[height= 2in]{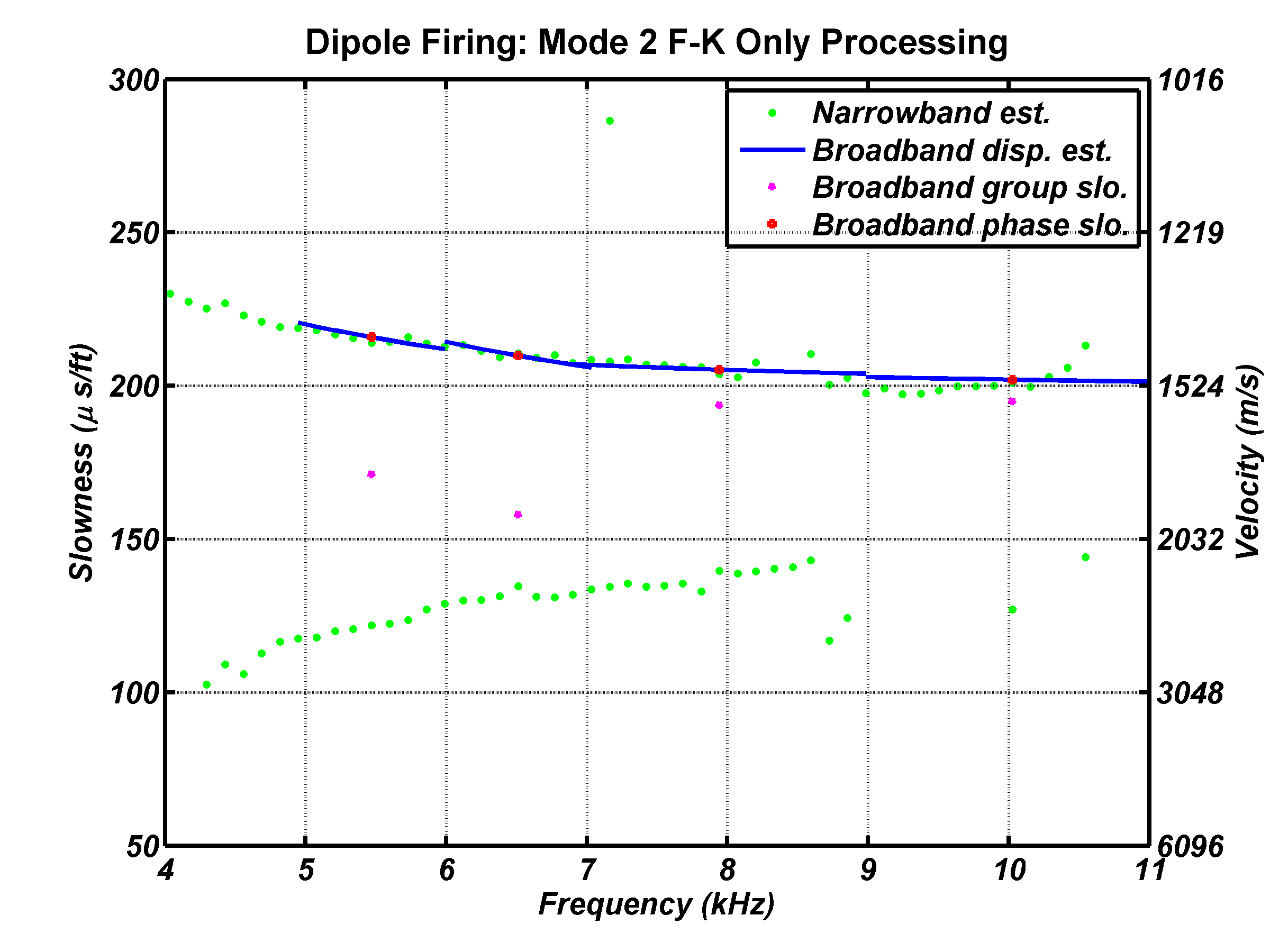}\\
      (a) &(b)\\
            \includegraphics[height= 2in]{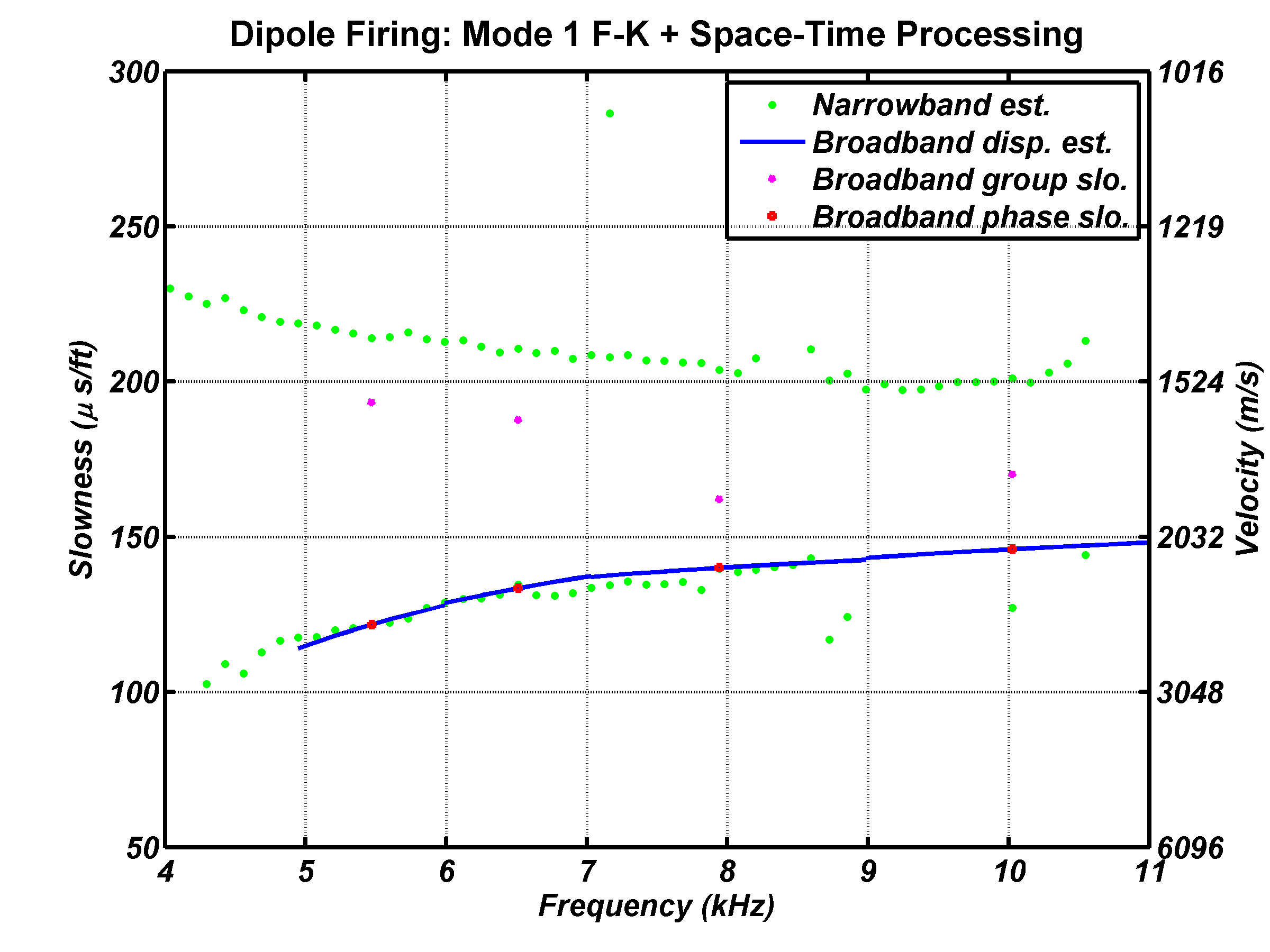} & \includegraphics[height= 2in]{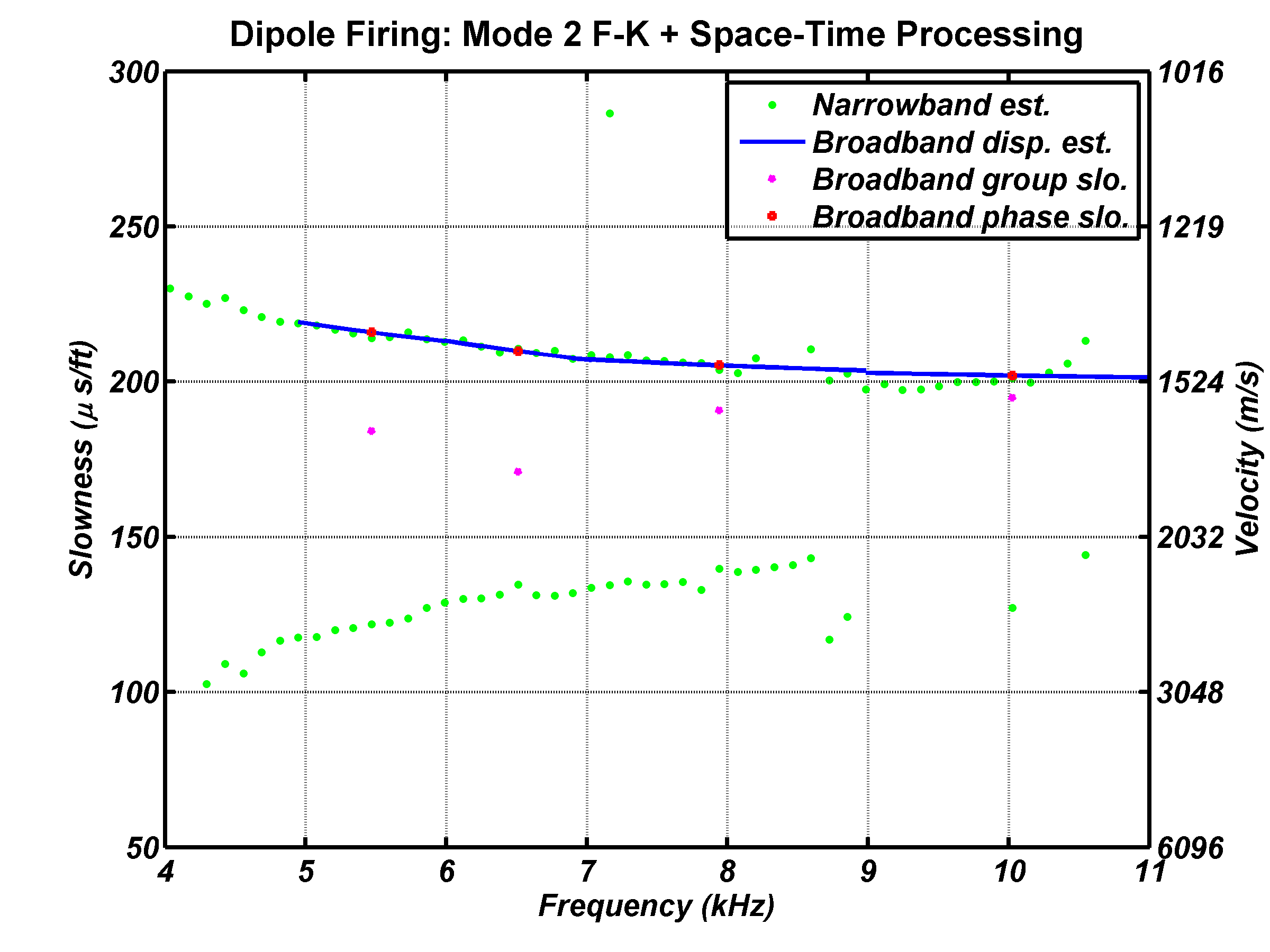}\\
(c)&(d)
      \end{tabular}}
  \caption{Dispersion extraction results of a LWD Dipole firing - Frame 2. The annotation and conclusions are similar to those in figure~\ref{fig:Dipole1}.}
  \label{fig:Dipole2}
\end{figure}

Figures~\ref{fig:Dipole1wave} \&  \ref{fig:Dipole2wave}  illustrate examples of received waveforms and spectra for dipole firings using an LWD tool in two different depth zones.    The formation is a fast formation and we observe the presence of the coupled modes that overlap in time and frequency.  Figures~\ref{fig:Dipole1} \& \ref{fig:Dipole2} display the results of slowness extraction using the proposed method on these data frames. As in the previous example, the use of the space time broadband processing results in significantly improved dispersion estimates addressing both the scatter of the narrowband processing and the artifacts induced by the broadband $f-k$ only processing.  In particular, we note the significant improvement in the estimate for the lower mode.  There may be scope for further improvement for the case in figure~\ref{fig:Dipole1} by performing one more iteration of refinement of the phase slowness using the updated group slowness estimates.

Due to the complexity of the LWD dipole acquisition, it is more common to use the monopole and quadrupole sonic acquisitions for LWD formation evaluation.   In those scenarios, under ideal conditions with a well centered LWD tool, these produce single dominant modes more amenable to conventional processing.  However in more complex cases resulting from tool eccentering or significant formation anisotropy, it is possible to obtain multiple overlapping modes which need to be extracted and analyzed for proper interpretation.

\begin{figure}
\centering \makebox[0in]{
    \begin{tabular}{cc}
    \includegraphics[height= 2in]{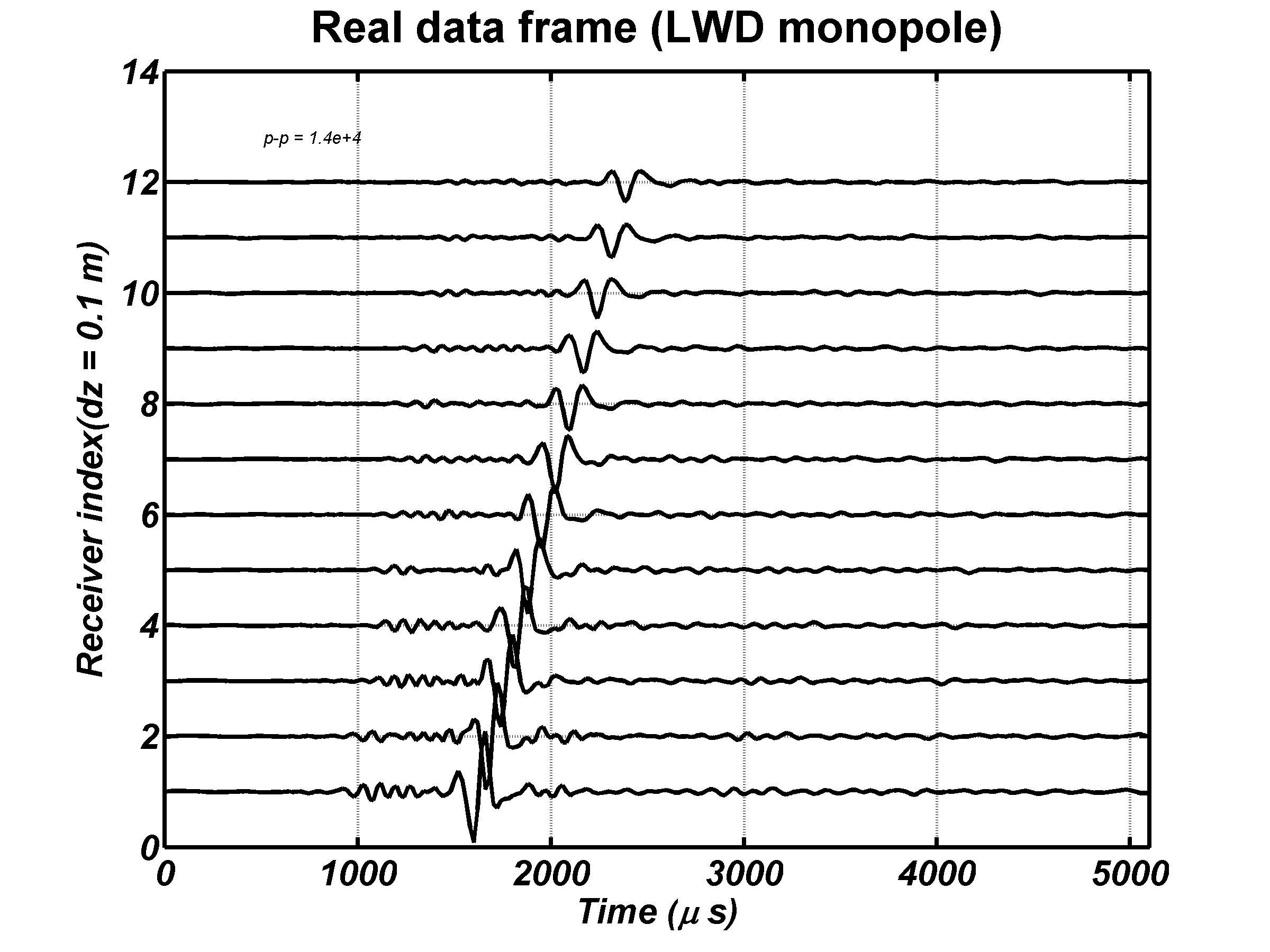} &
    \includegraphics[height= 2in]{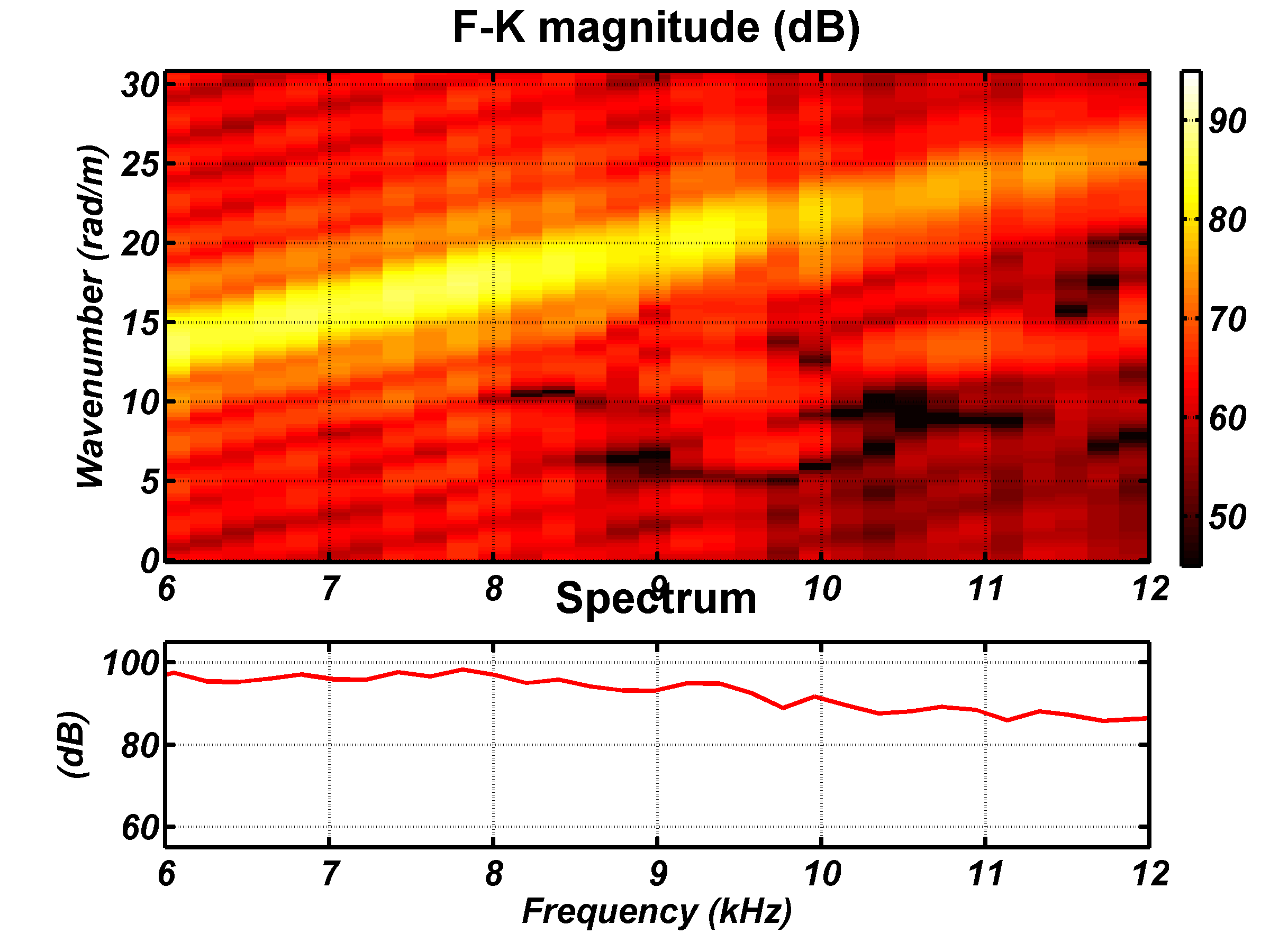}\\
  (a) & (b) 
   \end{tabular}}
  \caption{An example of monopole data  acquisition obtained with an LWD  sonic  tool showing plots of (a) waveform array of traces on the left and (b)  f-k and spectrum on the right.  Note that the second mode, expected in this section traversing a bed boundary, is considerably weaker than the first and is hidden in the sidelobes of the first on the f-k plot.}
\label{fig:MonopoleWave}
\end{figure}

\begin{figure}
\centering \makebox[0in]{
    \begin{tabular}{cc}
      \includegraphics[height= 2in]{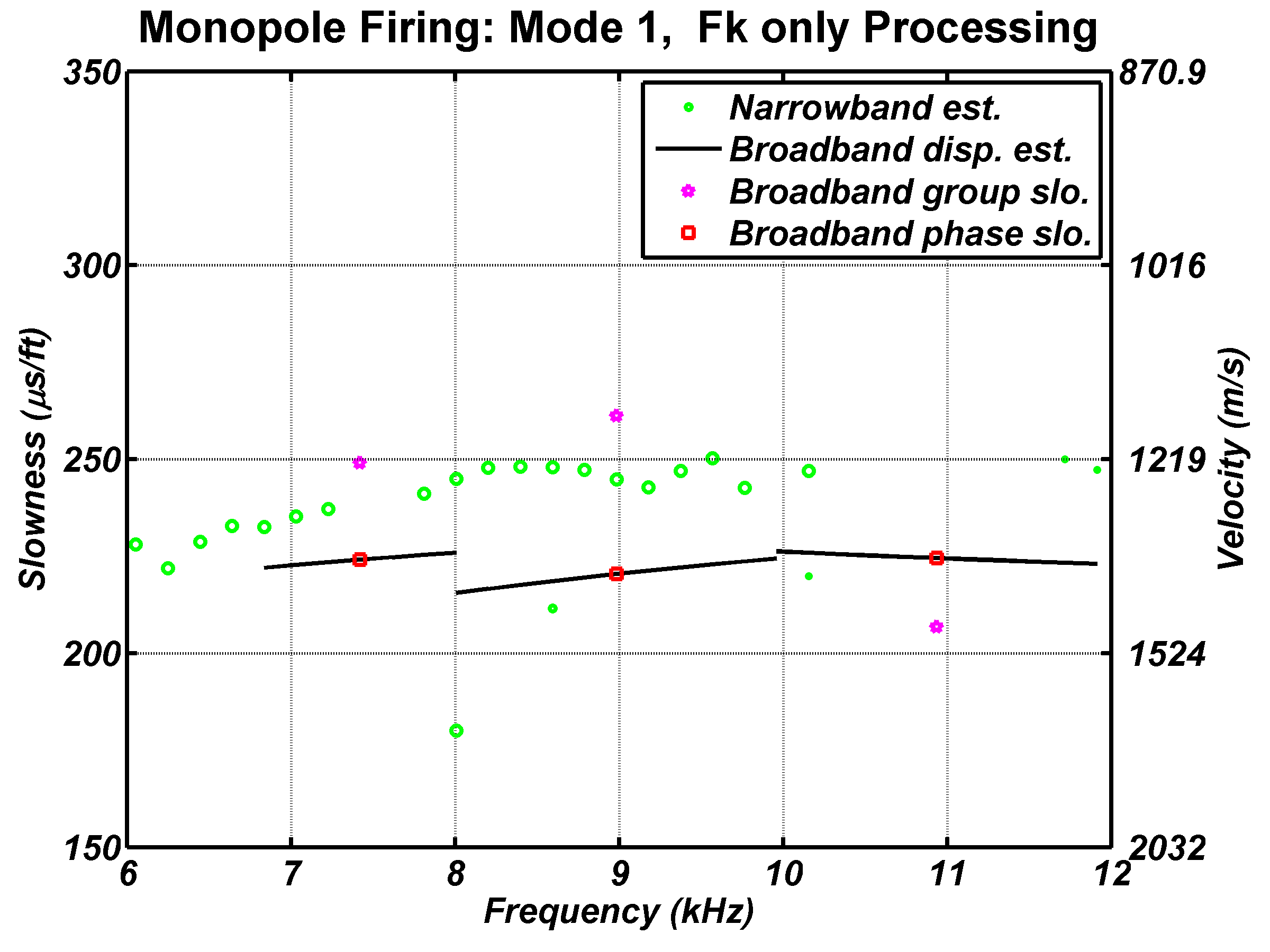} & \includegraphics[height= 2in]{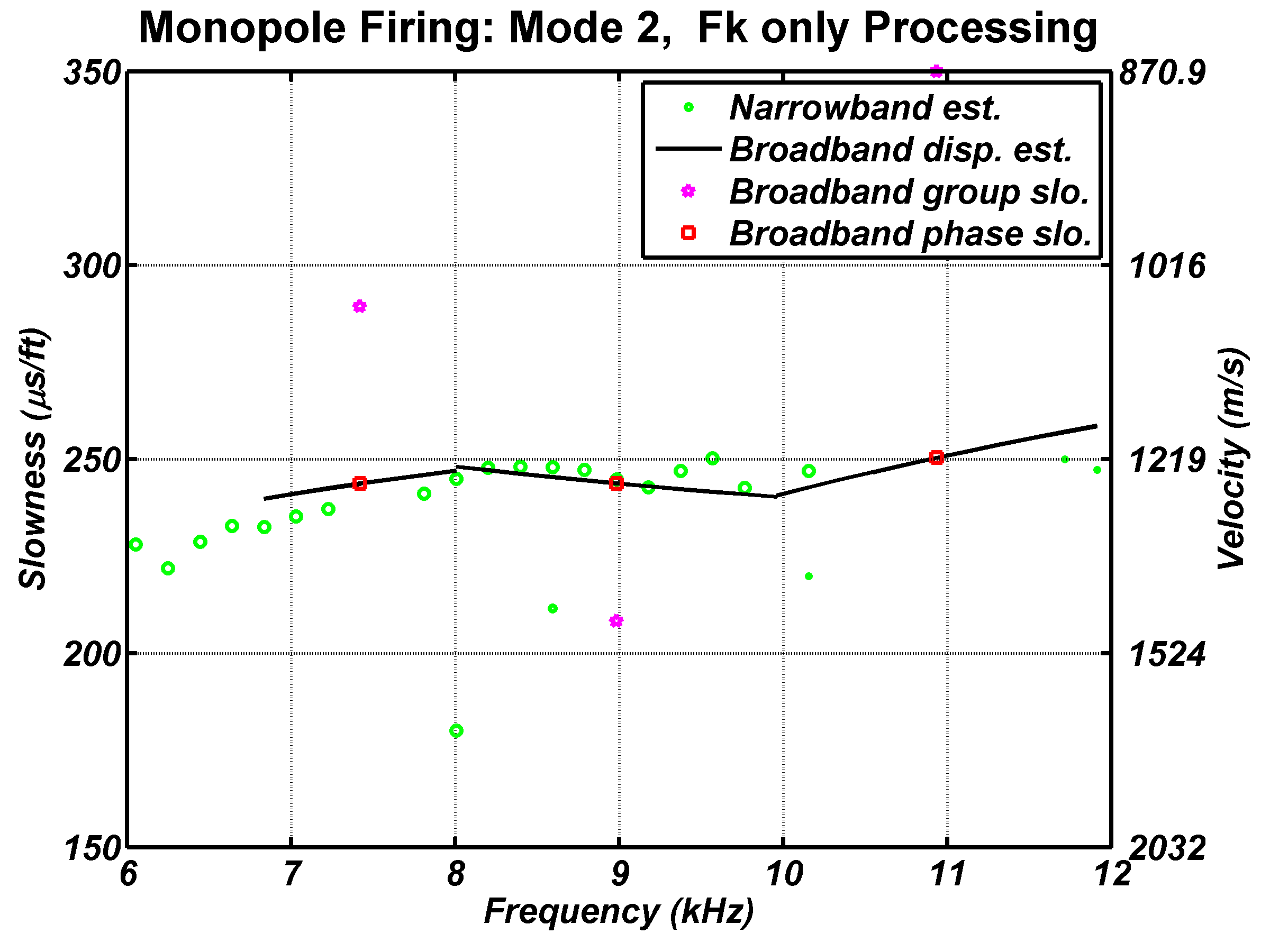}\\
      (a) &(b)\\
            \includegraphics[height= 2in]{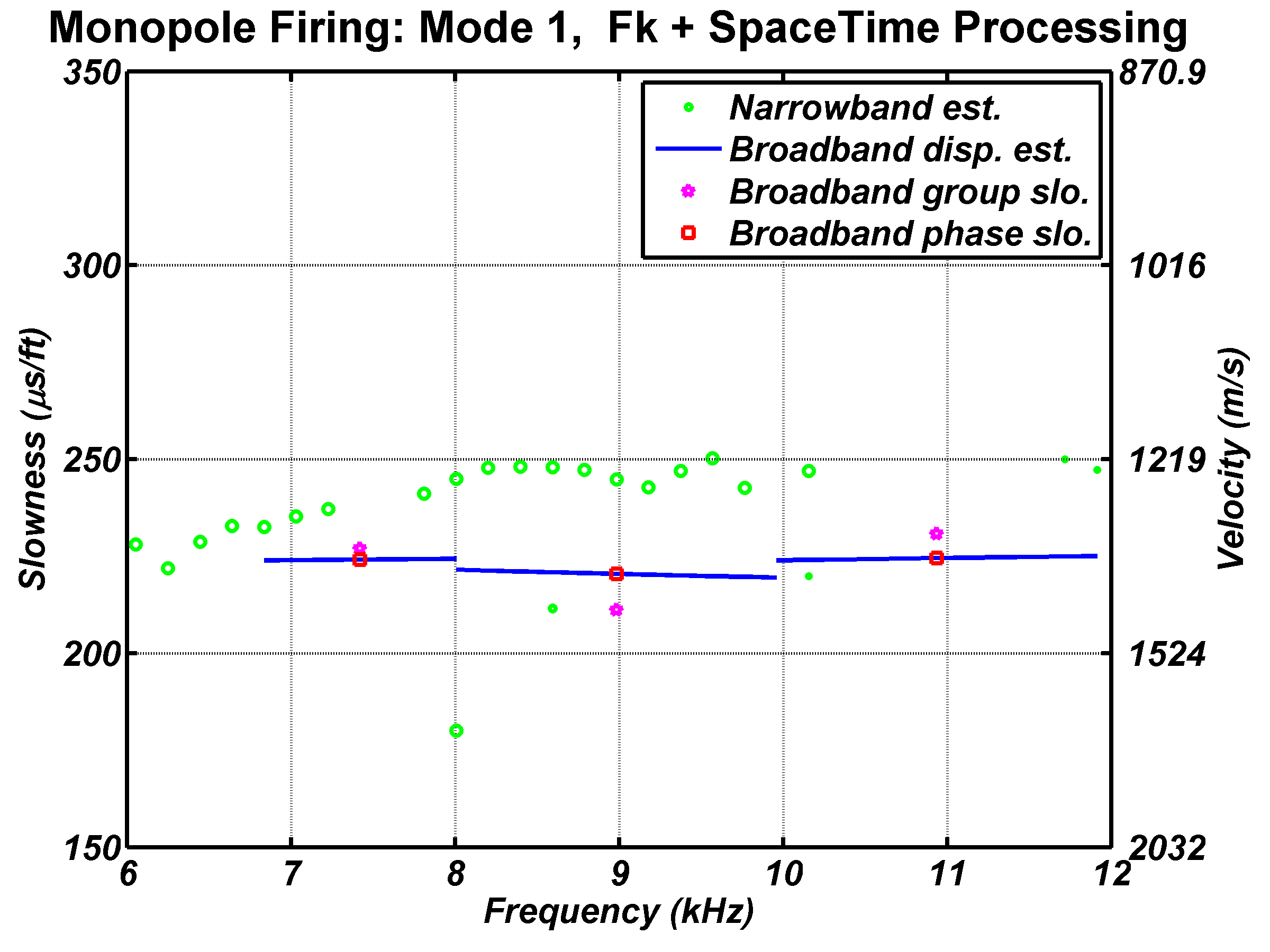} & \includegraphics[height= 2in]{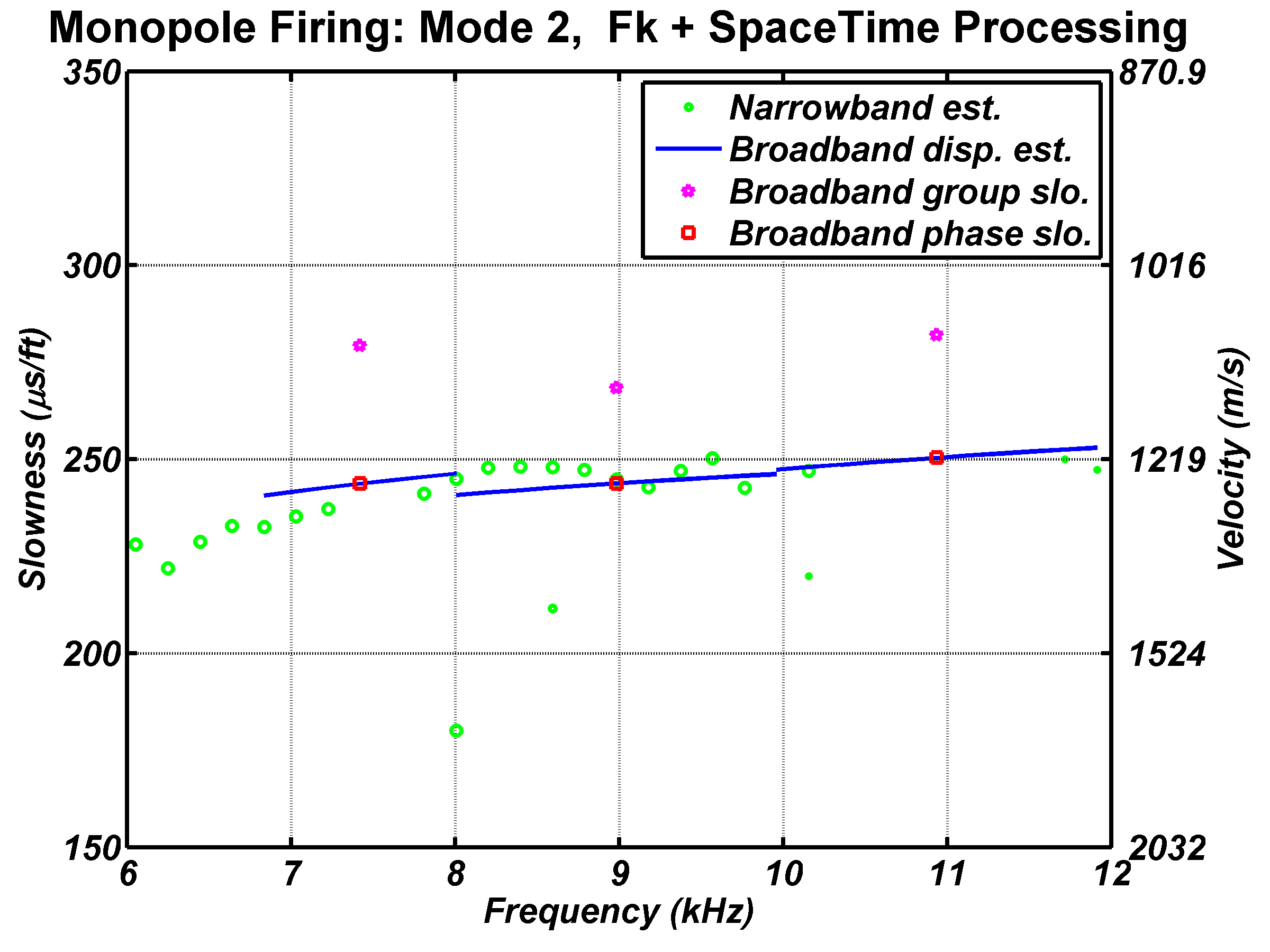}\\
(c)&(d)
      \end{tabular}}
  \caption{Dispersion extraction results on a frame of LWD monopole sonic data.  The annotation and conclusions are similar to those in the figure~\ref{fig:Dipole1}.  In this case, only the broadband approach is able to extract a dispersion curve for the weaker mode expected as explained in figure~\ref{fig:MonopoleWave}.}
  \label{fig:Monopole}
\end{figure}

We therefore apply our proposed approach to such a case for a monopole LWD acquisition,  illustrated in Figure~\ref{fig:MonopoleWave}, and display the result in figure~\ref{fig:Monopole}. We observe that the narrowband results exhibit considerable scatter making it hard to interpret and only a suggestion of a second mode.  The broadband approaches in contrast present considerably more stable answers with significant improvement seen with the space time approach, albeit with some scatter perhaps due to the challenging SNR environment.


\begin{figure}
\centering \makebox[0in]{
    \begin{tabular}{cc}
    \includegraphics[height= 2in]{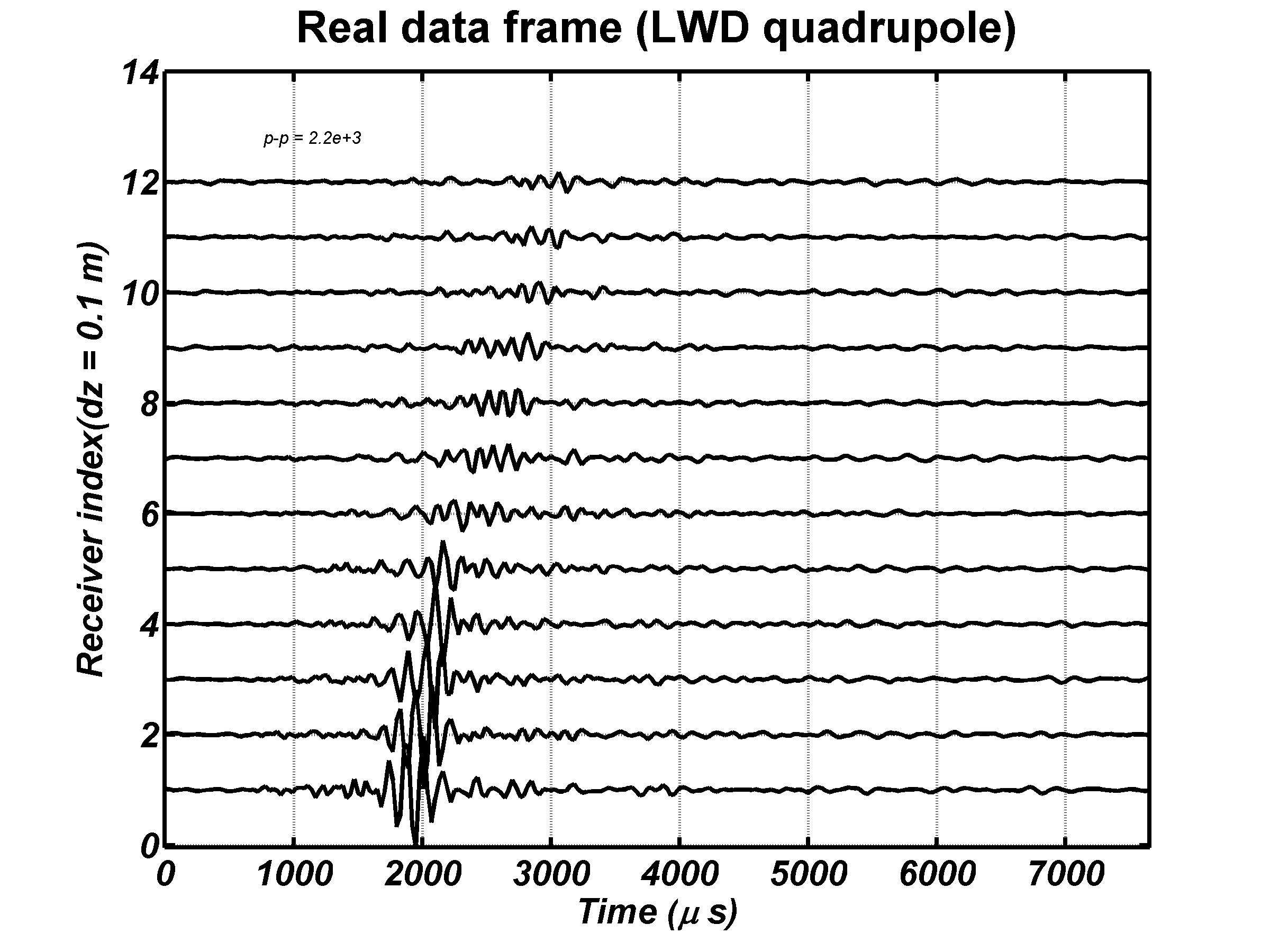} &
    \includegraphics[height= 2in]{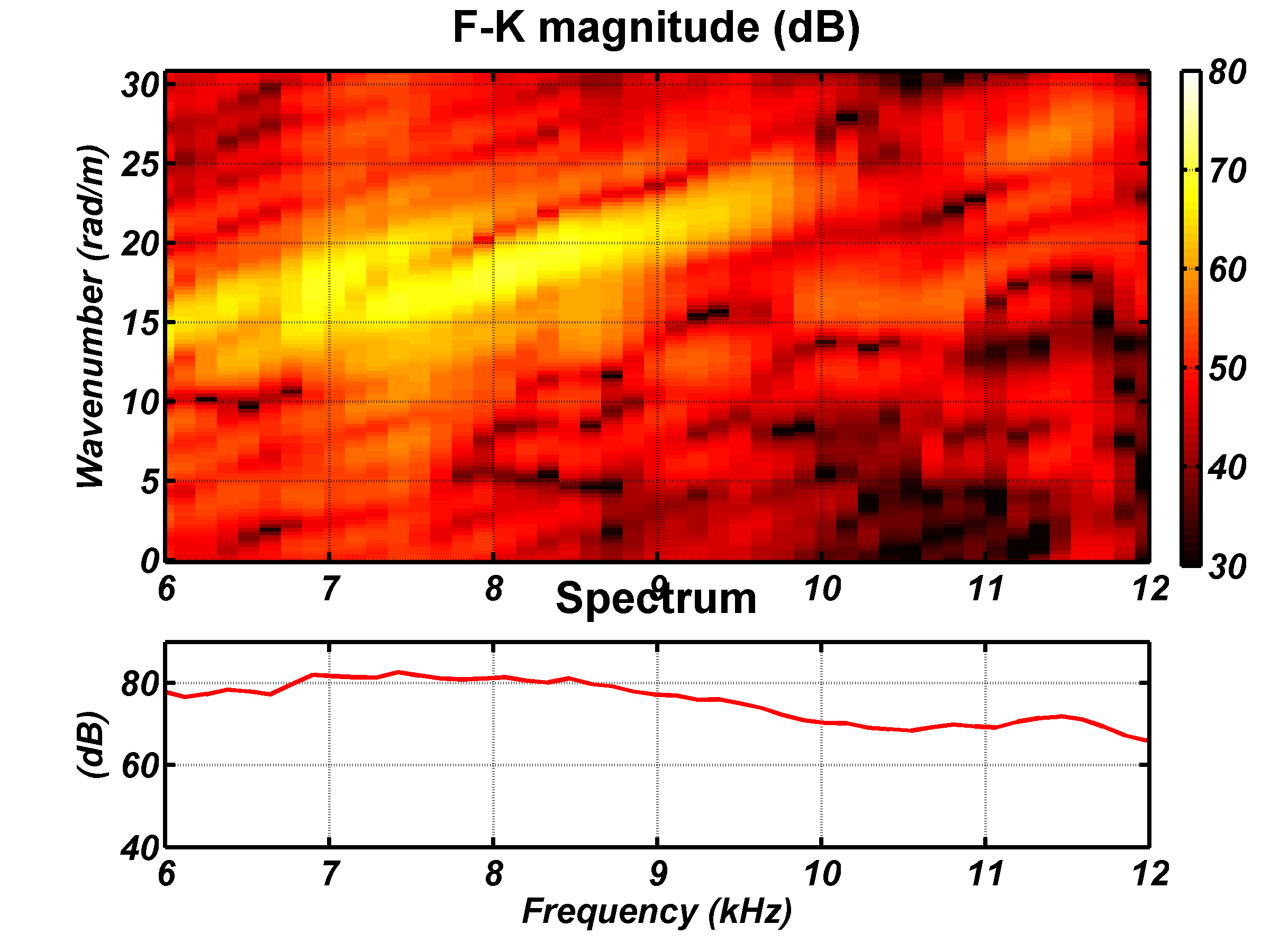} \\
  (a) & (b) 
   \end{tabular}}
  \caption{An example of quadrupole data  acquisition obtained with an LWD  sonic  tool showing (a) waveform array of traces on the left and (b)  f-k and spectrum on the right.  Observe the complexity of the wavetrain and closeness of the two modes on the f-k plot.}
\label{fig:QuadWave}
\end{figure}

\begin{figure}
\centering \makebox[0in]{
    \begin{tabular}{cc}
      \includegraphics[height= 2in]{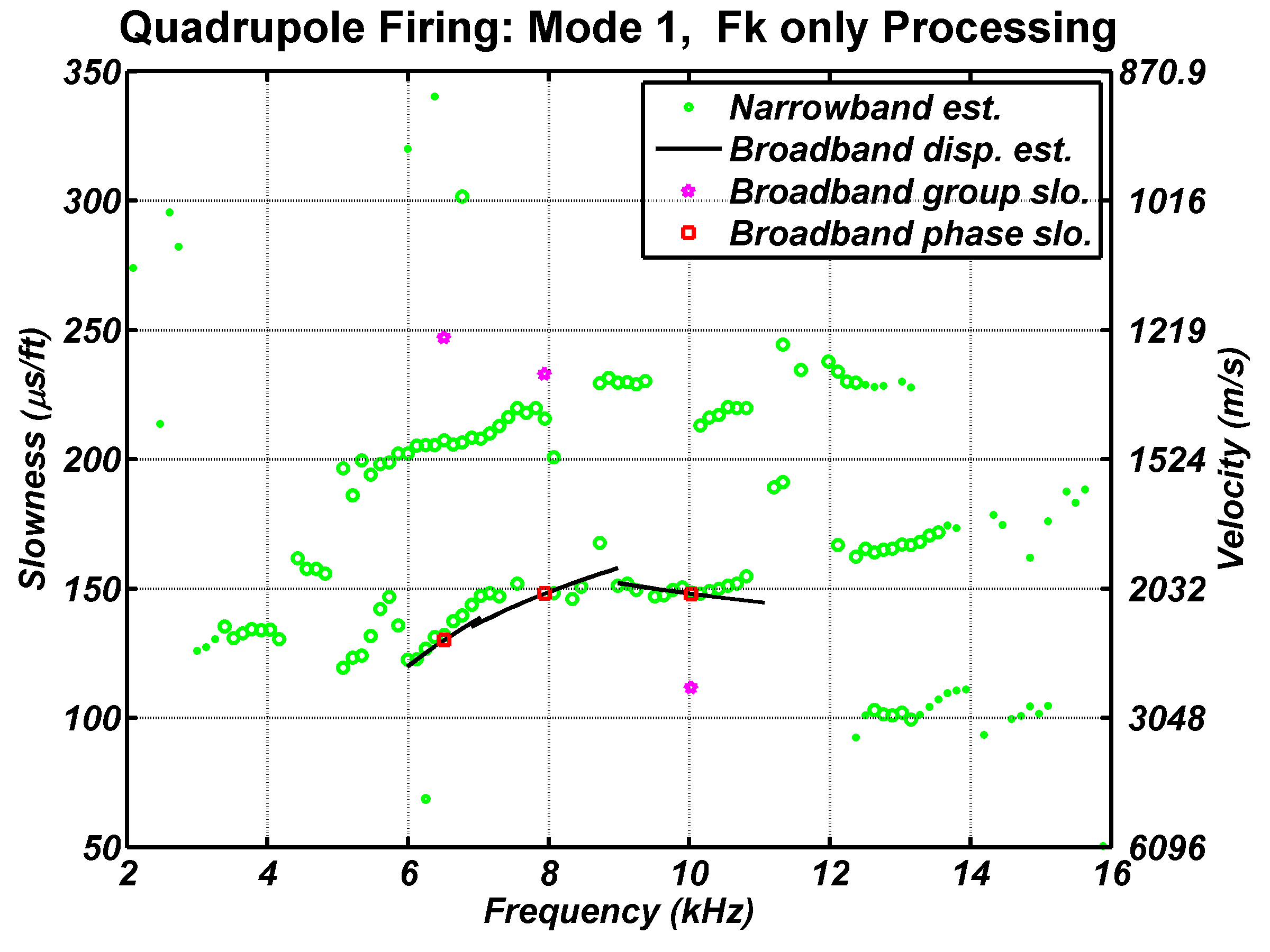} & \includegraphics[height= 2in]{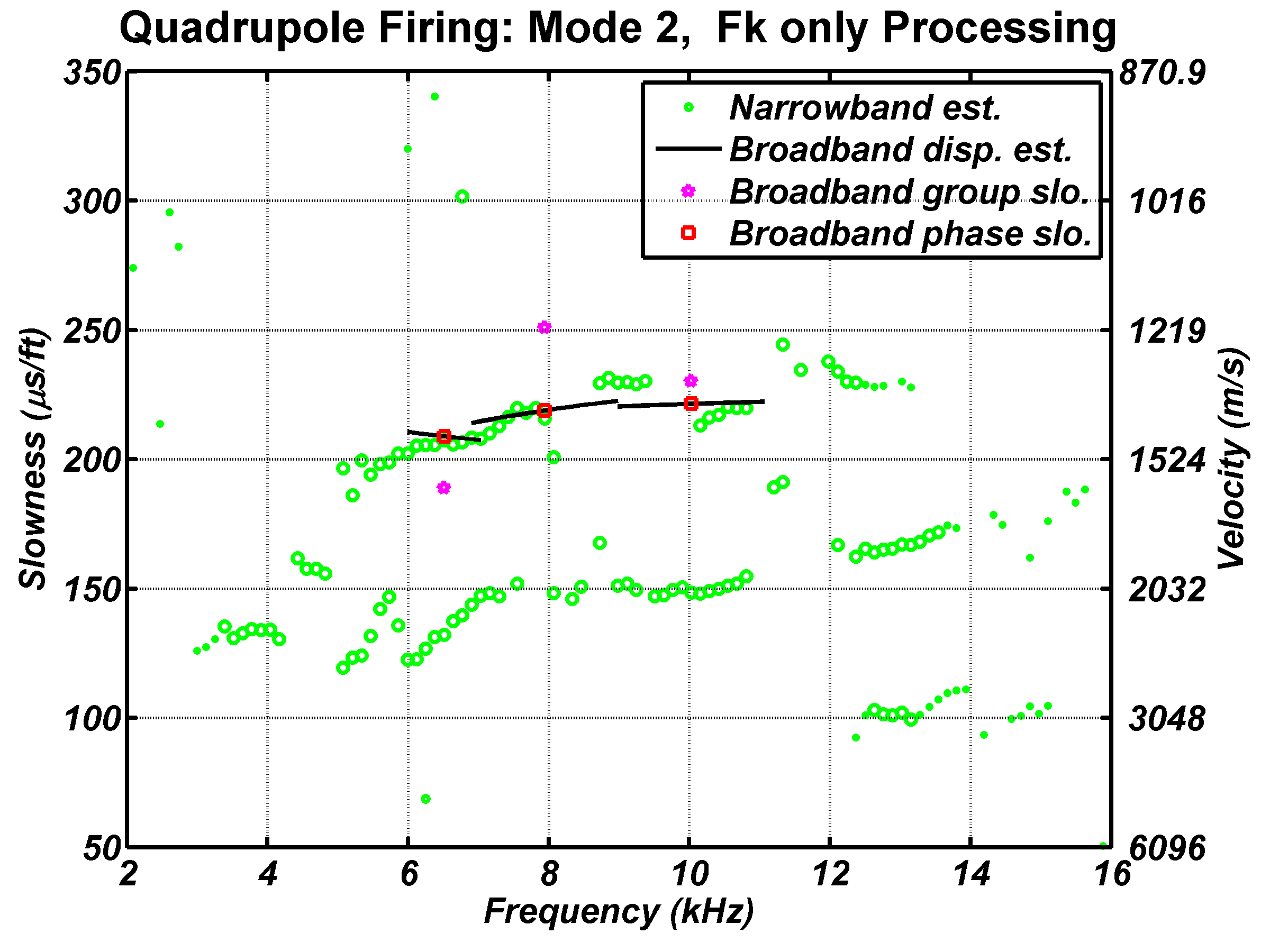}\\
      (a) &(b)\\
            \includegraphics[height= 2in]{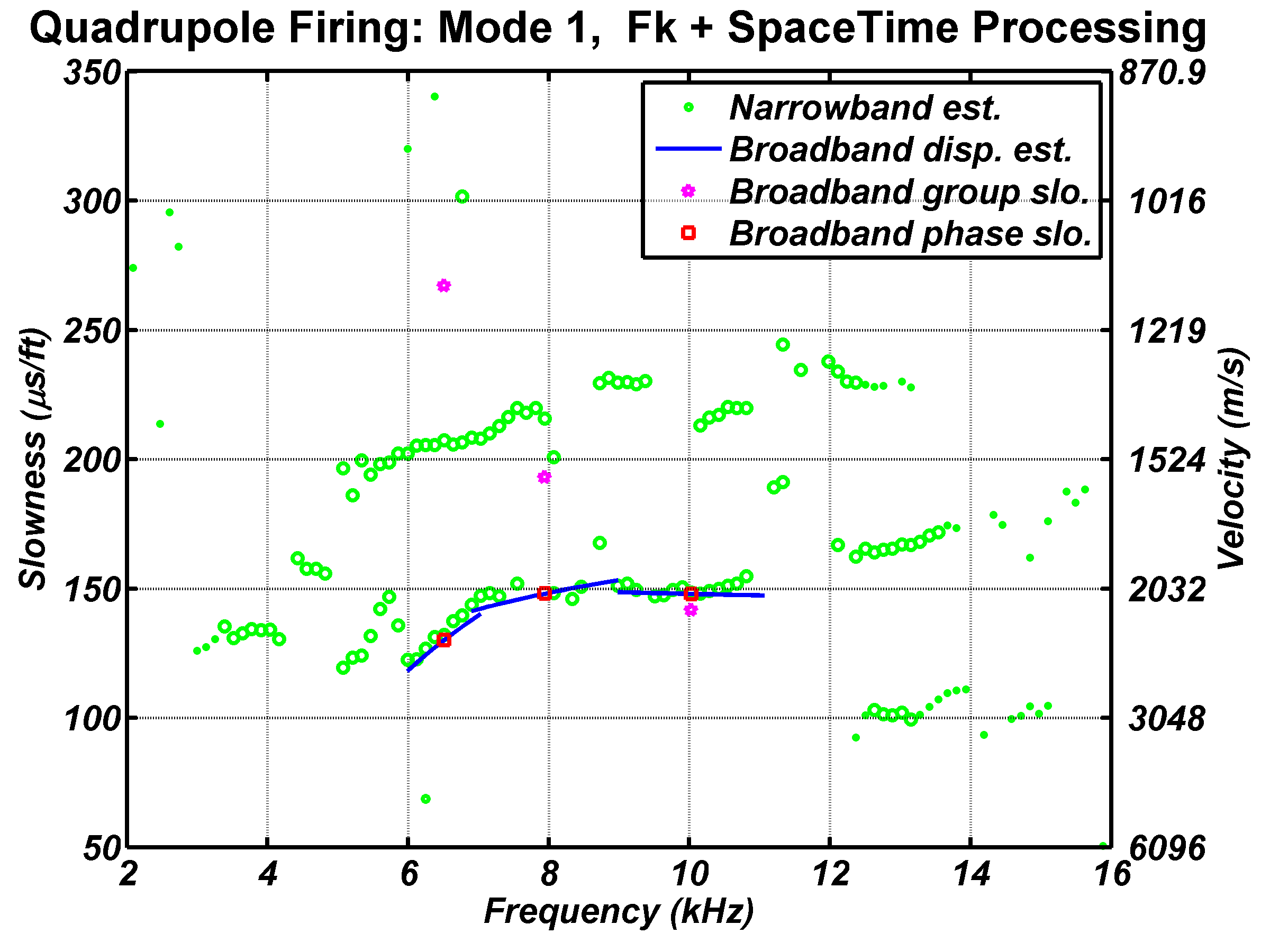} & \includegraphics[height= 2in]{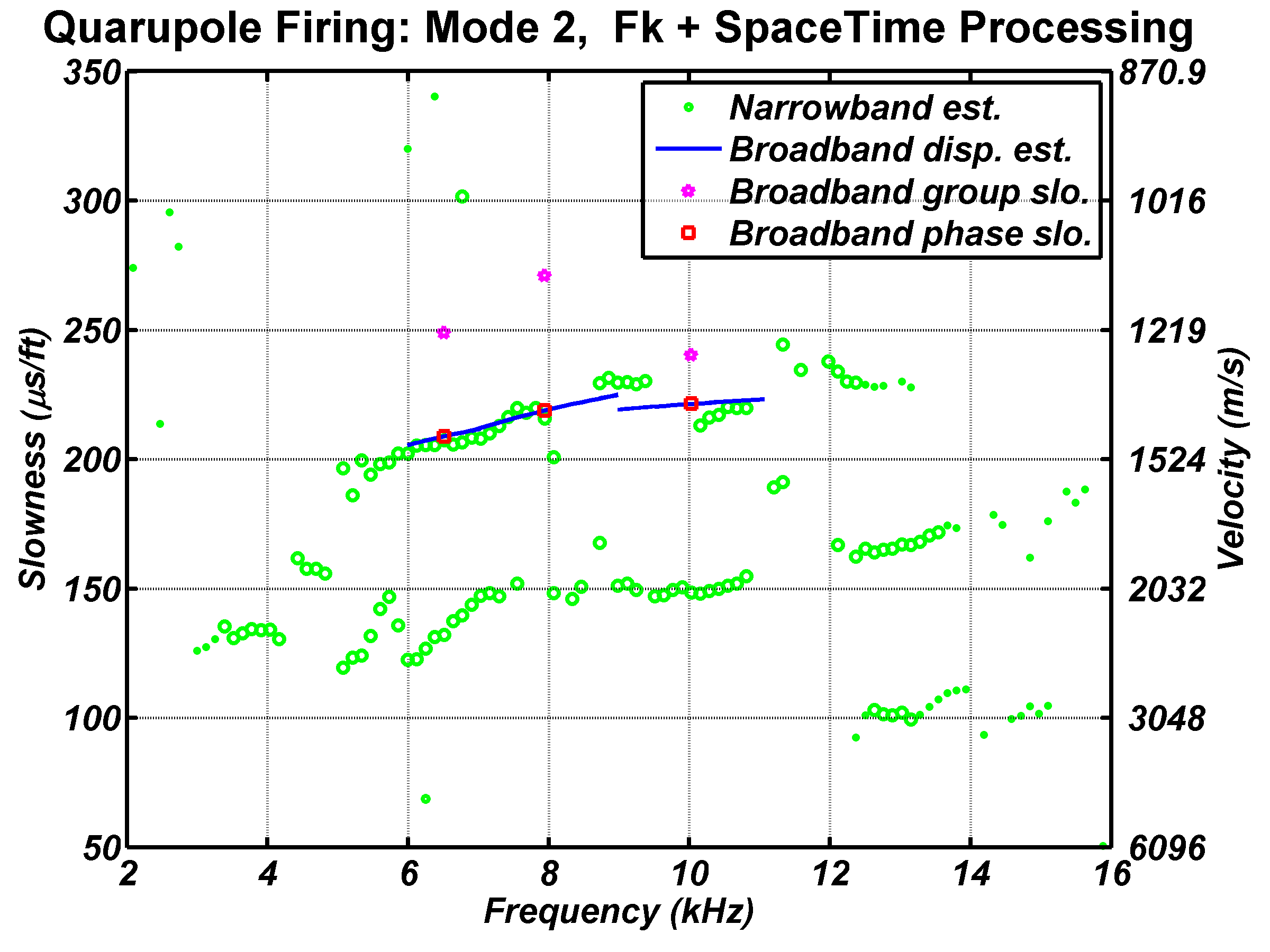}\\
(c)&(d)
      \end{tabular}}
  \caption{Dispersion extraction results on a frame of LWD quadrupole acquisition. The annotation and conclusions are similar to those in figure~\ref{fig:Dipole1}.}
  \label{fig:Quadrupole}
\end{figure}

Finally we look at a case with an LWD quadrupole acquisition illustrated in figure~\ref{fig:QuadWave}  and show the results in figure~\ref{fig:Quadrupole}.   Again the physical complexity results in the presence of two major overlapping modes whose dispersions need to be extracted for proper interpretation.  The narrowband approach again yields a significant scatter of unlabeled point estimates.  The broadband approach especially with the space-time processing results in much improved dispersion estimates that can be used for further analysis.

\section{Conclusion and future work}

In this paper we combined the broadband f-k domain processing with the space time processing to obtain robust estimates of
{\emph group and phase slowness} dispersion curves. Through application on real data for a variety of scenarios, we showed the performance of the method to handle multiple time-frequency overlapped modes. In particular significant improvements in the quality and consistency of the dispersion estimates was observed.  As a future work it may be possible to further enhance the performance by explicitly exploiting the continuity of the dispersion curves across disjoint frequency bands which is not currently being done. Nevertheless the significant improvements in the group slowness estimates with the broadband space time approach presented here facilitates the extraction of continuous dispersion curve estimates {\emph allowing for finer and more accurate interpretation of borehole acoustic data}.

\section{Appendix}
\label{sec:CRB}
\subsection{Cramer-Rao Bound for slowness estimation in f-k domain}

Before we go into the details of the derivation of the Cramer Rao Bound (CRB) we would like to introduce the following notation. For a matrix $\bA$ the transpose is denoted by $\bA^{T}$ and the conjugate is denoted by $\bA^*$. The conjugate transpose is then denoted by
$(\bA^{*})^T$. To this end recall that the Fisher Information Matrix $\mathbf{J} (\Theta)$ as a function of the parameters
$\Theta$ is given by,
\begin{align}
\label{eq:FIM1} [\mathbf{J}(\Theta)]_{i,j} = - \ex \left\{ \frac{\partial}{\partial \theta_i \partial \theta_j} \log
\prob(\bY|\Theta)\right\}
\end{align}

Then the CRB matrix is given by $\mathbf{J}^{-1}$. In the following we will derive general expressions for the various parts of
the Fisher Information Matrix for the following set-up. The observations are obtained according to the observation model,
\begin{align}
\bY = \bAt \bX + \bW
\end{align}
where we assume that the noise $\bW$ is i.i.d. complex Gaussian with zero mean and variance $\sigma^{2}$ in each dimension. In
our context $\bY \in \Complex^{L. N_f \times 1}$ is the Fourier transform of the array data in the band $F$, $\bAt$ denotes the
array response (exponential) dependent on the real parameters $\Theta$ that capture the phase and the group slowness of the modes
and $\bX \in \Complex^{M.N_f \times 1}$ is the mode spectrum corresponding to the $M$ modes.  We define the SNR as an overall SNR
given by
\begin{align}
\mbox{SNR} = \dfrac{||\bAt \bX||_{2}^{2}}{||\bW||_{2}^{2}}
\end{align}

Under this set-up the Log-Likelihood function $LL_{\Theta,\bX} = \log \prob(\bY|\Theta)$ as a function of $(\Theta,\bX)$ is given
by
\begin{align}
- LL_{\Theta,\bX} & = \frac{1}{ \sigma^{2}} (\bY^* - \bAt^* \bX^*)^T(\bY - \bAt \bX) + c \\
& =  \frac{1}{ \sigma^{2}} (\bY^*)^T \bY - (\bX^*)^T (\bAt^*)^T \bY - (\bY^*)^T \bAt \bX \nonumber\\
&\hspace{3 mm} + (\bX^*)^T(\bAt^*)^T \bAt \bX + c
\end{align}
where the constant term $c$ is only dependent on the noise variance $\sigma^2$ and the dimension of the problem. In order to
evaluate CRB the FIM has to be evaluated with respect to amplitudes $|\bX|$ and the real parameters in  $\Theta$. Note that the
derivative in equation \ref{eq:FIM1} involve complex quantities. The common approach in this case is to represent each quantity
by its real and imaginary parts and take derivatives with  respect to these. But more compact expressions are obtained using
\emph{sesquilinear} convention. Such a form is suitable for real functions of complex quantities, e.g. $Q$ and $Q^*$ (say) that
can be expressed as functions of $Q \pm Q^*$ or $(Q^*)^T Q$.   For taking the derivatives we then simply regard $Q$ and $Q^*$ as
independent variables and take derivatives with respect to them. One can then go back to the real quantities by applying a
suitable transformation.

Therefore in order to evaluate the CRB we evaluate the Fisher Information Matrix (FIM) with variables $\bX, \bX^*, \Theta$ and
where $\Theta$ is the vector of parameters that we are trying to estimate. In order to apply the sesquilinear convention we note
that for a real function $f(\bAt,\bAt^{*})$
\begin{align}
\frac{\partial}{\partial_{\theta}} f(\bAt, \bAt^{*})  = & Tr \left\{\left(\frac{\partial}{\partial_{\bAt}} f(\bAt,
\bAt^{*})\right)^{T} \frac{\partial \bAt}{\partial_{\theta}}\right. \nonumber\\
& + \left.\left(\frac{\partial}{\partial_{\bAt^*}} f(\bAt,
\bAt^{*})\right)^{T} \frac{\partial \bAt^{*}}{\partial_{\theta}} \right\}
\end{align}
where $Tr(.)$ is the Trace operation. In the following we will use the following standard conventions used for matrix calculus -
For matrices $A,X,B$
\begin{align}
\frac{\partial}{\partial X} Tr(A X B)  &= A^T B^T \\
\frac{\partial}{\partial X} Tr(AX^T B) & = B A
\end{align}
Using the above relation we have,
\begin{align}
-\sigma^2 \frac{\partial}{\partial \bAt} LL_{\Theta} =  \left( - \bY^* \bX^{T}  +   \bAt^{*} \bX^* \bX^{T}\right)
\end{align}
\begin{align}
-\sigma^2 \frac{\partial}{\partial \bAt^{*}} LL_{\Theta} =   \left(- \bY (\bX^{*})^{T} + \bAt\bX (\bX^*)^{T} \right)
\end{align}
This implies the following.
\begin{align}
-\sigma^2 \frac{\partial}{\partial \theta} LL_{\Theta,\bX}  = & Tr \left\{ \left( -\bX (\bY^{*})^{T} + \bX (\bAt^* \bX^*)^T
\right)\frac{\partial \bAt}{\partial_{\Theta}} \right. \nonumber \\
& +  \left.\left( - \bX^* \bY^T + \bX^*(\bAt\bX)^T  \right)\frac{\partial
\bAt^{*}}{\partial_{\theta}} \right\}
\end{align}

After taking the derivative again with respect to $\theta$ we note that the terms involving double derivatives in $\bAt$ and
$\bAt^*$ evaluate to zero under the expectation operation. Therefore we have,
\begin{align}
- \sigma^{2} \ex \left[\frac{\partial^2}{\partial \theta^2} LL_{\Theta,\bX}\right]  = & Tr \left\{ \bX (\bX^*)^T
\left(\frac{\partial \bAt^{*}}{\partial_{\theta}}\right)^T \frac{\partial \bAt}{\partial_{\theta}} \right. \nonumber \\
& + \left. \bX^*\bX^T
\left(\frac{\partial \bAt}{\partial_{\theta}}\right)^T \frac{\partial \bAt^{*}}{\partial_{\theta}} \right\}
\end{align}
Now note that
\begin{align}
-\sigma^2 \frac{\partial}{\partial \bX} LL_{\Theta,\bX} = - \bAt^T \bY^* + \bAt^T \bAt^* \bX^*
\end{align}
This implies,
\begin{align}
-\sigma^2 \frac{\partial}{\partial \bX} \frac{\partial}{\partial \theta} LL_{\Theta,\bX} =&  (\bY^*)^T - (\bX^*\bAt^*)^T
\frac{\partial}{\partial \theta}\bAt \nonumber \\
& + \bAt^T  (\frac {\partial \bAt^*}{\partial \theta})^T (\bX^*)^T
\end{align}
The above  implies that
\begin{align}
-\sigma^2 \ex \left[\frac{\partial}{\partial \bX} \frac{\partial}{\partial \theta} LL_{\Theta,\bX} \right] =  \bAt^T
(\frac{\partial \bAt^*}{\partial \theta})^T (\bX^*)^T
\end{align}

Using the above formulas it is easy to show that for all $\theta_i, \theta_j \in \Theta$,
\begin{align}
- \sigma^2 \ex \frac{\partial}{\partial \theta_i \theta_j} LL_{\Theta,\bX} = & Tr\left\{ \bX (\bX^*)^T \frac{\partial
(\bAt^*)^T}{\partial \theta_i}\frac{\partial \bAt}{\partial \theta_j} \right. \nonumber \\
& + \left.\bX^* (\bX)^T \frac{\partial (\bAt)^T}{\partial
\theta_i}\frac{\partial \bAt^*}{\partial \theta_j} \right\}
\end{align}
and for all $\theta_i \in \Theta$
\begin{align}
-\sigma^2 \ex \frac{\partial}{\partial \bX^*}\frac{\partial}{ \partial \theta_i } LL_{\Theta,\bX}  = (\bAt^*)^T \frac{\partial
\bAt}{\partial \theta_i } \bX
\end{align}
\begin{align}
-\sigma^2 \ex \frac{\partial }{\partial \bX}\frac{\partial}{ \partial \theta_i } LL_{\Theta,\bX}  = (\bAt)^T \frac{\partial
\bAt^*}{\partial \theta_i } \bX^*
\end{align}

\begin{align}
-\sigma^2 \frac{\partial }{\partial \bX^*}\frac{\partial}{\partial \bX} LL_{\Theta,\bX} & = \bAt^T\bAt^*\\
-\sigma^2 \frac{\partial}{\partial \bX}\frac{\partial}{\partial \bX^*} LL_{\Theta,\bX}& = (\bAt^*)^T\bAt
\end{align}

\subsection{Fisher Information for slowness estimation}

We will use the general expressions derived above for finding the CRB for slowness estimation for a 2-mode problem. To this end
note that
\begin{align}
\Theta = [ \theta_1, \theta_2, \theta_3, \theta_4 ] = [ k_1, k_{1}^{'}, k_2, k_{2}^{'} ]
\end{align}

Also let $\Theta_1  = [\theta_1, \theta_2] = [k_1(f_0), k_{1}^{'}(f_0) ]$ and  $\Theta_2 = [\theta_3, \theta_4]  = [k_2(f_0),
k_{2}^{'}(f_0) ]$ are the dispersion parameters for the two modes. Following the formulation used before we define  in this case
the matrix $\bAt$ is given by Equation \ref{eq:Atheta}.
\begin{figure*}
\begin{align}
\label{eq:Atheta}
\bAt = \begin{bmatrix} \bv_{\Theta_1}(f_1) &  &  & &  \bv_{\Theta_2}(f_1) &  &  &\\
                        & \bv_{\Theta_1}(f_2) &  &  &  &  \bv_{\Theta_2}(f_2)&  & \\
                        & & \ddots &  &  &  &   \ddots &\\
                        & & & \bv_{\Theta_1}(f_{N_f}) &  &  &  &  \bv_{\Theta_2}(f_{N_f})\end{bmatrix}
\end{align}
\end{figure*}
where
\begin{align}
\mathbf{v}_{\Theta_m}(f) = \begin{bmatrix}
   e^{-i2\pi (k_m + k_{m}^{'}(f - f_0))(z_1-z_0) } \\
  e^{-i2\pi (k_m + k_{m}^{'}(f - f_0)) (z_2 - z_0)}\\
  \vdots \\
  e^{-i2\pi (k_m+ k_{m}^{'}(f - f_0))(z_L - z_0)}
\end{bmatrix}
\end{align}
$ m = 1,2 , f \in \left\{f_1,f_2,...,f_{N_f}\right\}$. Then
\begin{align}
- \sigma^2 \ex \frac{\partial}{\partial \theta_i \theta_j} LL_{\Theta,\bX}   = & Tr\left\{ \bX (\bX^*)^T \frac{\partial
(\bAt^*)^T}{\partial \theta_i}\frac{\partial \bAt}{\partial
\theta_j} \right. \nonumber \\
& \left. + \bX^* (\bX)^T \frac{\partial (\bAt)^T}{\partial \theta_i}\frac{\partial \bAt^*}{\partial \theta_j} \right\}
\end{align}
The above expression is equal to,
\begin{align}
 2 \;Tr \; \mbox{Real}\; \left\{ \sum_{f_i} \bX_1(f_i) \bX_{2}^{*}(f_i) \left[ \frac{\partial}{\partial \theta_i}
(\bB_{\Theta}^{*}(f_i))^T \frac{\partial}{\partial \theta_j} \bB_ {\Theta}(f_i) \right]\right\}
\end{align}
where  $ \bB_{\Theta}(f_1) = [\bv_{\Theta_1}(f_1),\bv_{\Theta_2}(f_1) ]$  and  $\bX_1, \bX_2 \in \Complex^{N_f \times 1}$ are
the coefficients of the mode spectrum for the two modes. For the cross terms it can be shown that
\begin{align}
(\bAt)^T \frac{\partial \bAt^*}{\partial \theta_i} \bX^* = \begin{bmatrix} \begin{bmatrix} \bv_{\Theta_1}^{T} \frac{\partial
\bv_{\Theta_1}^{*}}{\partial \theta_i} \bX_{1}^{*}
(f_1) + \bv_{\Theta_1}^{T} \frac{\partial \bv_{\Theta_2}^{*}}{\partial \theta_i} \bX_{2}^{*}(f_1) \\
 \bv_{\Theta_2}^{T} \frac{\partial \bv_{\Theta_1}^{*}}{\partial \theta_i} \bX_{1}^{*}(f_1) + \bv_{\Theta_2}^{T} \frac{\partial \bv_{\Theta_2}^{*}}{\partial \theta_i} \bX_{2}^{*}
(f_1)
\end{bmatrix} \\
\vdots \\
\vdots \end{bmatrix}
\end{align}
and similar expressions can be obtained for the other term  $ (\bAt^*)^T \frac{\partial \bAt}{\partial \theta_i } \bX$.  Now note
that for the particular choice of $z_0$ to be the median of the array locations $z_1,z_2, ...z_L$ the terms,
\begin{align}
\bv_{\Theta_1}^{T} \frac{\partial \bv_{\Theta_1}^{*}}{\partial \theta_i} \bX_{1}^{*}(f_1) = 0 \\
\bv_{\Theta_2}^{T} \frac{\partial \bv_{\Theta_2}^{*}}{\partial \theta_i} \bX_{2}^{*}(f_1)  = 0
\end{align}
This means that the slowness estimates decouple from the amplitude estimates for each mode. However the interference terms
remain. For given values of the phase and group slowness and the mode spectrum one can then calculate the CRB from the general
expression for the FIM obtained above.   In order to obtain the CRB in terms of amplitudes and phase one simply applies the
following transformation
\begin{align}
\mathbf{K}^T  FIM  \mathbf{K}
\end{align}
where
\begin{align}
\mathbf{K} = \begin{bmatrix}  \mbox{diag} \left(\frac{\bX}{|\bX|}\right) &   \mbox{diag} \left(\frac{\bX^*}{|\bX|}\right) & 0 \\
                          i  \mbox{diag} (\bX)  & - i  \mbox{diag}( \bX^*) &  0 \\
                          0 & 0 & \mathbf{I} \end{bmatrix}
\end{align}
to the FIM as obtained above using the sesquilinear analysis before taking the inverse. In the above equation
$\mbox{diag}(\cdot)$ of a vector is a diagonal matrix with the diagonal being composed of the elements from the vector.  Below we
will derive simple error bounds to time location estimates using the error bounds on the phase location estimates.

\subsection{Obtaining error bounds for time location estimates}
\label{sec:CRB_time_loc}

One can use the error bounds on the phase estimates obtained from the CRB analysis above to obtain error bounds on the time
location estimates. Indeed in principle one may further parameterize the set-up in terms of the linear phase of the mode spectrum
but here we will take an alternate approach. We will essentially use the error bound on the slope for the Least Squares Fit
obtained from erroneous data , \cite{Acton66} to bound the error on the time location estimate. From CRB let the error variance
in the phase estimates of the mode spectrum of a mode be given by - $e_1,...,e_{N_f}$. Then the variance in the estimation of
slope at points $f_1,...,f_{N_f}$ is given by,
\begin{align}
\mbox{Var(slope)} = \frac{es}{es_{ff}}
\end{align}
where
\begin{align}
es^2  & = \frac{1}{N_f - 2} \sum_{i=1}^{N_f} e_{i}^{2} \\
es_{ff}  & = \sum_{i=1}^{N_f}(f_i  - \bar{f})^2, \;\;\; \bar{f} = \frac{1}{N_f} \sum_{i=1}^{N_f} f_{i}
\end{align}

\bibliographystyle{IEEEtran}
\bibliography{IEEE_TSP_Initial_Sub}

\end{document}